\documentclass[structabstract]{aa}  
\usepackage{natbib}	
\usepackage{txfonts,epsfig,graphicx,url,twoopt}
\usepackage[breaklinks=true]{hyperref} 
\hypersetup{colorlinks=true,citecolor=blue}
\bibpunct{(}{)}{;}{a}{}{,}   
\usepackage[usenames,dvipsnames]{color}

\begin{document}

\title{Sequential planet formation in the HD~100546 protoplanetary disk?}

   \author{P.~Pinilla\inst{1}, T.~Birnstiel\inst{2} and C.~Walsh\inst{1}}
   \institute{Leiden Observatory, Leiden University, P.O. Box 9513, 2300RA Leiden, The Netherlands\\
              \email{pinilla@strw.leidenuniv.nl}
              \and
              Harvard-Smithsonian Center for Astrophysics, 60 Garden Street, Cambridge, MA 02138, USA
              }
   \date{Accepted 25 May 2015}

 
\abstract
{The disk around the Herbig~Ae star, HD~100546, shows structures that suggest the presence of 
two companions in the disk at $\sim10$ and $\sim70$~AU. The outer companion seems to be in the act of formation.}
{Our aims are to provide constraints on the age of the planets in HD~100546 and to explore the potential evidence 
for sequential planet formation in transition disks such as HD~100546.}
{We compare the recent resolved continuum observations of the disk around HD~100546 with the results of dust evolution simulations 
using an analytical prescription for the shapes of gaps carved by massive planets.}
{An inner pressure bump must have been present since early in the disk lifetime to have good agreement between the 
dust evolution models and the continuum observations of HD~100546. 
This pressure bump may have resulted from the presence of a 
very massive planet ($\sim20~M_{\rm{Jup}}$), which formed early in the inner disk ($r\sim$10~AU). 
If only this single planet exists, the disk is likely to be old, comparable to the stellar age ($\sim$5-10~Myr). 
Another possible explanation is an additional massive planet in the outer disk ($r\sim70$~AU):  
either a low-mass outer planet ($\lesssim5~M_{\rm{Jup}}$) injected at early times, or a higher mass outer 
planet ($\gtrsim15~M_{\rm{Jup}}$) formed very recently, traps the right amount of dust in pressure bumps to reproduce the observations. 
In the latter case, the disk could be much younger ($\sim3.0$~Myr).}
{In the case in which two massive companions are embedded in the disk around HD~100546, as suggested in the literature, 
the outer companion could be at least $\gtrsim$2.5~Myr younger than the inner companion.}
  

\maketitle

\section{Introduction}     \label{introduction}
Transition disks display different interesting structures such as dust depleted cavities \citep[e.g.][]{brown2009, andrews2011}, 
azimuthal asymmetries \citep[e.g][]{isella2013, casassus2013, marel2013, perez2014}, spiral arms \citep[e.g.][]{fukagawa2004, muto2012, grady2013}, 
and spatial segregation of small and large particles \citep[e.g.][]{follette2013, garufi2013, zhang2014, hashimoto2014}. 
These structures suggest that these disks host a massive planet or multiple planets. 
Observations of planet candidates in transition disks \citep[e.g][]{huelamo2011, kraus2012, biller2012, quanz2013} 
have further supported this idea. 
Nonetheless, other mechanisms such as photoevaporation may also play an important role and explain some of the 
observed structures \citep[e.g.][]{alexander2006, owen2012, rosotti2013}. 

Observations of the disk around the Herbig~Ae star HD~100546 have indicated the presence of two potential companions. 
By modelling the [O~I] $6300~\AA$ emission line obtained with VLT/UVES in HD~100546, \cite{acke2006} suggested the 
presence of a massive planet ($\gtrsim20~M_{\rm Jup}$) at $\sim6.5$~AU distance from the star. 
Using VLTI/AMBER at H- and K-band, \cite{tatulli2011} proposed a less massive planet ($\sim1-8~M_{\rm Jup}$) 
in the inner disk. 
However, using MIDI/VLT observations and by constraining the curvature of the disk wall of the inner cavity, 
\cite{mulders2013} found a lower limit for the mass of the inner companion. 
Taking temperature changes when a planet opens a gap into account, \cite{mulders2013} conclude that the mass of a potential planet located at $\sim~10$AU is $20-30~M_{\rm Jup}$. 
On the other hand, high-contrast imaging with VLT/NACO shows signatures of a massive planet 
($\sim15~M_{\rm Jup}$) at $\sim~70$AU, which may be its formation stage \citep{quanz2013, quanz2014}. 
\cite{currie2014} reported Gemini/NICI thermal infrared data and detected the outer protoplanet at the 
same location and brightness as that found by \cite{quanz2013}. 
This emission seems to have an extended structure which may come from a circumplanetary disk. 
In addition, spiral arms have been observed in scattered light images 
\citep[e.g.][]{grady2001, ardila2007, boccaletti2013, avenhaus2014}, 
which may be related to the presence of these planets.

Previous observations of protoplanetary disks have revealed that the disk radial extent can be much larger 
for the molecular gas than for the millimetre-sized particles. 
This is the case for TW~Hya, whose CO emission extends up to $\sim$215AU, 
while the mm grains extend to only $\sim$60AU from the star \citep{andrews2012}. 
\cite{birnstiel2014} suggest that radial drift may be responsible for the different radial extents of the gas and large dust grains. 
However, the disk around HD~100546 appears to be a special case: the radial extent of the gas is $\sim400$AU, 
while most of the emission at millimetre wavelengths comes from a narrow ring concentrated at $\sim$~26AU, 
with a width of $\sim$21~AU. 
In addition, a much fainter ring of emission (a factor of $\sim$100 lower) comes from the outer disk, 
which is centred at $\sim190$~AU, with a width of $\sim$~75 AU \citep{walsh2014}.  
This double-ring emission is consistent with the two-planet scenario \citep{walsh2014}.   
Ring-like emission in transition disks can be explained by the dust evolution,  
which occurs when a single massive planet or multiple planets interact with the disk 
\citep[e.g.][]{pinilla2012, pinilla2014b, zhu2014}. 
An alternative explanation for multiple rings is magneto-rotational instability in the outer 
regions of disks \citep[e.g.][]{pinilla2012b, flock2014}. 
In protoplanetary disks, particles migrate inward because of the sub-Keplerian motion of the gas 
\citep[e.g.][] {Weidenschilling1977, Nakagawa1981, tanga1996, birnstiel2010}. 
When a massive planet opens a gap, particles stop their inward migration due to the positive 
pressure gradient at the outer edge of that gap. Dust grains accumulate and grow in this 
preferential region known as a pressure trap, whose location and structure depend on 
disk viscosity,  the mass and location of the planet \citep[e.g.][]{rice2006}.

In this work, we investigate the influence of the two suggested planets in HD~100546 on the dust 
distribution in the disk, and compare the resulting predicted continuum emission with the most 
recent mm observations. 
By computing dust evolution models for different parameters, we aim to put constraints on different 
planet properties, and to address the following questions: can dust evolution, without any planet(s) 
in the disk, explain the two-component emission observed with ALMA? What happens when a single 
inner companion is assumed?  What if two companions are assumed? 
To answer these questions, we cover a large parameter space, which includes the mass, location, and age 
of the planets, and disk viscosity.
\footnote{By age, we mean the time since the injection of the planet into the simulations}. 

For HD~100546, proper hydrodynamical simulations for the planet-disk interaction are  
highly computationally demanding because of the large radial separation between the two planet candidates. 
Since we aim to study several cases with different planet properties, we instead use analytical solutions 
to model the shapes of gaps in the disk. 
We use the \cite{crida2006} prescription for the width, and the \cite{fung2014} prescription for the depth. 
In Sect.~\ref{models}, we explain the assumptions for the carved gaps, the potential caveats, and the 
connection with dust evolution models.  
In Sect.~\ref{results} we present the results of the dust evolution models, the computed visibilities  
at different wavelengths ($\lambda=[0.87 , 1.0, 3.0, 7.0]$~mm), and the comparison with ALMA and 
ATCA observations of HD~100546. 
Sections~\ref{discussion} and \ref{conclusion} are the discussion and conclusions of this work.\\

\begin{figure*}
 \centering
 \tabcolsep=0.05cm 
   \begin{tabular}{cc}   
   	\includegraphics[width=8.5cm]{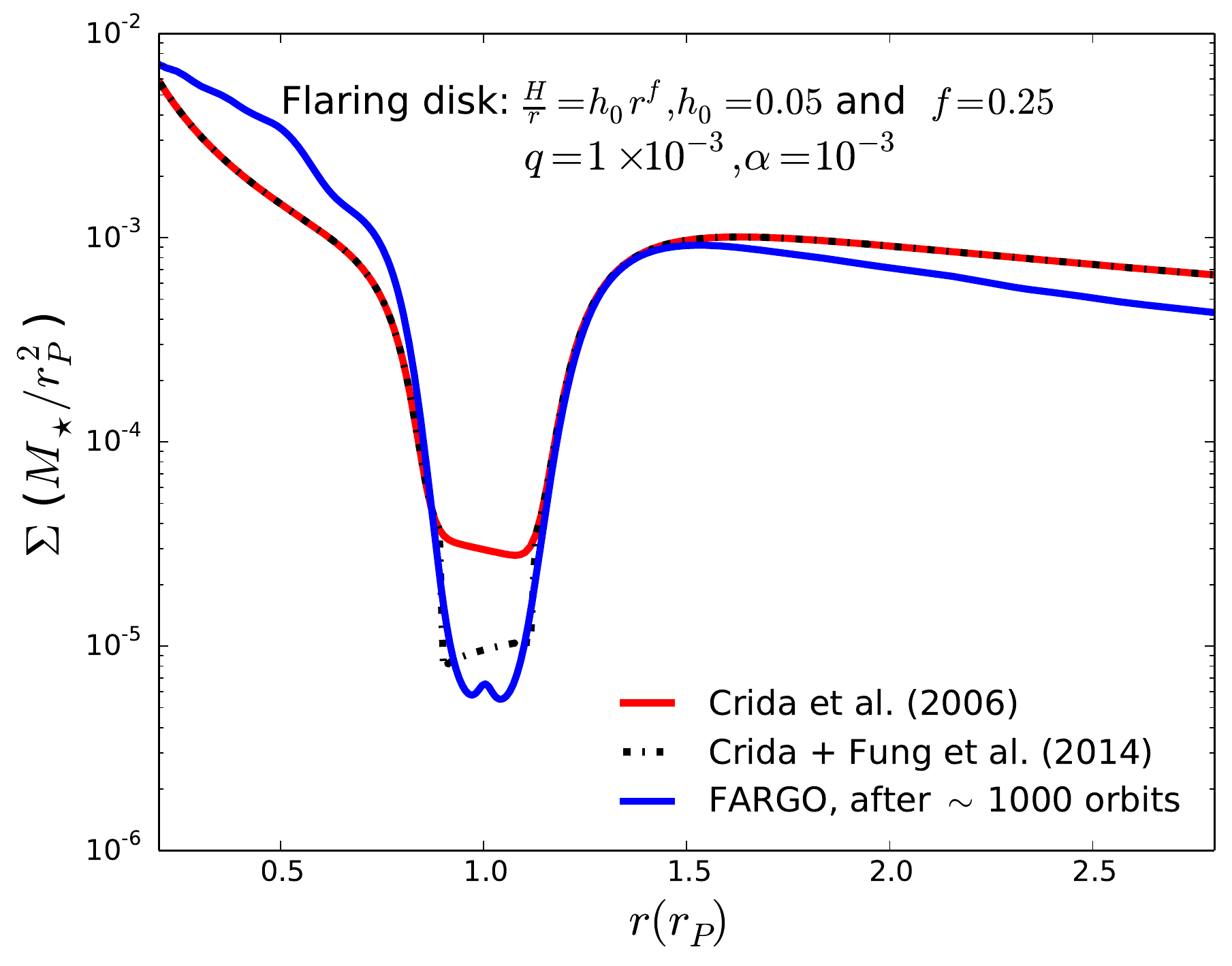}&
	\includegraphics[width=8.5cm]{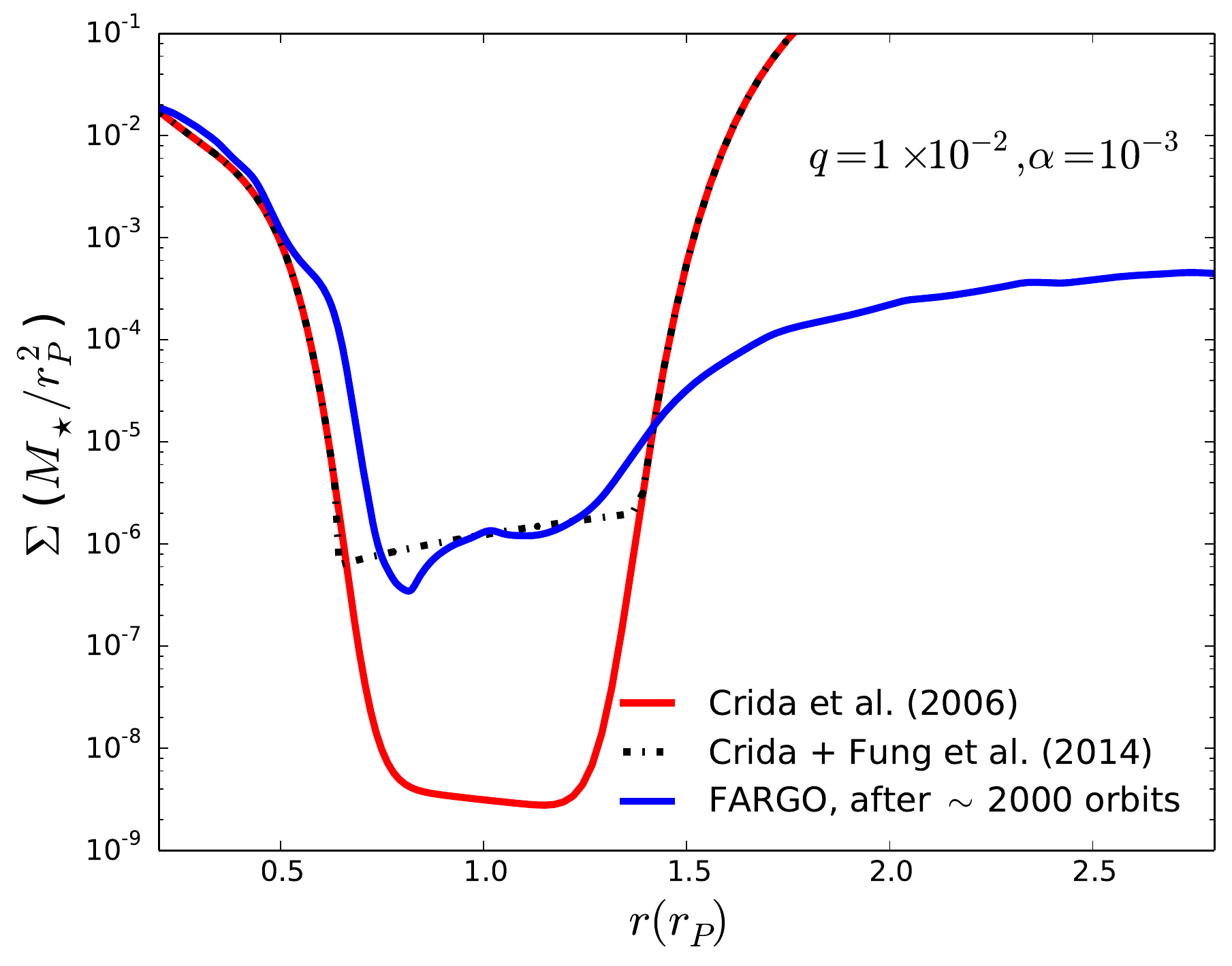}
   \end{tabular}
   \caption{Comparison between the analytical gap prescriptions and hydro-simulations when a massive planet opens a gap. 
   The disk is assumed to be a flared disk, i.e. $H/r=h_0r^f$, with $h_0=0.05$ the aspect ratio at the position of the planet, 
   and $f$ flaring index equal to 0.25. The planet-to-star mass ratio is $q=1\times10^{-3}$ (\emph{left panel}) 
   and $q=1\times{10^{-2}}$ (\emph{right panel}), and the planet position is $r_p=1$. 
   The disk viscosity is assumed to be $\alpha=10^{-3}$ for both cases. }
   \label{comparison_hydro}
\end{figure*}

\section{Models and set-up}     \label{models}

In this section, the analytical approximation for the shape of the gaps carved by planets is described, 
together with the dust evolution models, and the computation of the visibilities. 
To test this analytical approach and the validity of the resulting dust density distributions, different benchmark cases are considered. 
 
\subsection{Carved gaps} \label{carved_gaps}

Because of the large separation between the potential planets in HD~100546 ($\sim10$AU and $\sim70$AU), 
hydrodynamical simulations are computationally expensive. 
High grid resolution is needed close to the planet locations, and the inner and outer radial boundaries 
should be far enough from the planet positions to avoid unphysical results. 
Because our interest is focussed on the influence of the gas surface density carved by planets on the radial dust evolution, 
we use an approximation of the radial shape of the gaps in protoplanetary disks rather than full hydrodynamical simulations.  
We assume the analytical results presented in \cite{crida2006} and \cite{fung2014}. 
\cite{crida2006} proposed an equilibrium profile considering the viscous torque ($t_\nu$), 
gravitational torque ($t_g$), and the torque removed by pressure supported waves or pressure torque ($t_P$). 
This analytical profile therefore satisfies that $t_\nu + t_g + t_P =0$. 
In an $\alpha-$disk \citep{shakura1973}, where disk viscosity is parametrised as $\nu=\alpha c_s^2/\Omega$, 
the equilibrium solution for the gas surface density ($\Sigma$) when a planet located at $r_p$ is interacting with 
the disk \citep[see][Eqs.~11, 13, and 14]{crida2006} is

\begin{equation}
\left(\frac{r_H}{\Sigma}\frac{d\Sigma}{dr}\right)=\frac{t_g-\frac{3}{4}\alpha c_s^2}{\left(\frac{H}{r}\right)^2 r r_p \Omega_p^2 a'' + \frac{3}{2}\alpha c_s^2 \frac{r}{r_H}},
\label{crida_solution}
\end{equation}

\noindent with $r_H$ being the Hill radius of the planet ($r_H=r_p(q/3)^{1/3}$, 
where $q$ is the planet-stellar mass ratio $M_P/M_\star$, and $r_p$ the planet orbital radii), 
$\Omega_p$ the angular orbital velocity of the planet, $H/r$ the aspect ratio, $c_s$ the sound speed, 
and $a''$ a dimensionless function given by

\begin{equation}
a''\left(\frac{(r-r_p)}{r_H}\right)=\frac{1}{8}\left|\frac{(r-r_p)}{r_H}\right|^{-1.2}+200\left|\frac{(r-r_p)}{r_H}\right|^{-10}.
\label{crida_solution2}
\end{equation}

This dimensionless function $a''$ is an ansatz from 2D vertically isothermal simulations. 
When Eq.~(\ref{crida_solution}) is solved, a boundary condition needs to be imposed, which in our case 
is assumed to be the unperturbed density. 
Because the equilibrium solution of Eq.~(\ref{crida_solution}) assumes the gravitational and pressure torques 
to be null close to the planet ($-2r_H<r-r_p<2r_H$), the depth of the gap is not perfectly constrained. 

An analytical solution for the depth of the gap close to the planet is difficult to calculate because of the 
strong tidal gravitational field of the planet in this region. 
Using two independent codes (\texttt{PEnGUIn} and  \texttt{ZEUS90}), 
\cite{fung2014} provided an empirical relation of the depth of the gap carved by a non-migrating giant planet 
in a locally isothermal disk. 
They consider a large parameter space for the planet-stellar mass ratio $q$, viscosity $\alpha$, and aspect ratio $H/r$, 
finding very similar results with both codes. 

We imitate the gas gaps carved by massive planets by solving Eq~(\ref{crida_solution}) and 
correcting the depth of the gap with the empirical scalings \citep[equations 12 and 14 from][]{fung2014}, 
between $-2r_H<r-r_p<2r_H$. 
Both the width and the depth of the gaps are essential for the final dust distributions in disks. The width 
determines the location of the pressure maximum at the outer edge of the gap and therefore the location of the 
peak of the millimetre emission \citep[e.g.][]{pinilla2012}.  The depth is also important because of the possible dust 
filtration \citep[e.g][]{rice2006, zhu2012}.

To compare the results from these approximations, Fig.~\ref{comparison_hydro} displays the 
comparison of the gas surface density between hydrodynamical simulations done with \texttt{FARGO} \citep{masset2000}, 
and the gap shape obtained using the solution by \cite{crida2006} and corrected by the empirical relations from \cite{fung2014}. 
For this comparison, it is considered that a massive planet, with $q=1\times10^{-3}$ (left panel) and $q=1\times{10^{-2}}$  (right panel),  
opens a gap in a flared disk, i.e. $H/r=h_0r^f$, with $h_0=0.05$ being the aspect ratio at the position of the planet, 
and $f$ a flaring index equal to 0.25.  
The disk viscosity is assumed to be $\alpha=10^{-3}$. 
From the hydrodynamical simulations, the gas surface density is azimuthally averaged after the disk reaches a steady-state,  
1000 and 2000 planet orbits for $q=1\times10^{-3}$ and $q=1\times10^{-2}$ respectively.   
As shown in Fig.~\ref{comparison_hydro}, this approximation for the shape of the gaps is very good for the  
case of $q=1\times10^{-3}$.  
In the case of $q=1\times10^{-2}$, the width is under-predicted compared to the azimuthally-averaged profile from the 
hydrodynamical simulations. 
When a very massive planet $q\gtrsim5\times10^{-3}$ interacts with the disk, the formed gaps become eccentric \citep{kley2006}, 
making the radial gap profile wider and less steep. 
The location of the pressure bump at the outer edge of the gap may change from $7-8~R_H$ to $9-10~R_H$ 
for a non-eccentric to an eccentric gap. 
Thus, in the case of transition disks, to reproduce the peak of the millimetre emission at a certain distance from the star 
using this analytical approach, 
we give an approximate value for the planet location rather than a specific position when  $q\gtrsim5\times10^{-3}$. 
Since the match between the hydrodynamical simulations and the analytical approach only works for the gap shape, 
and potential differences become significant far from the gap location (Fig.~\ref{comparison_hydro}), 
we assume the gas surface density to follow the unperturbed density far from the location of the planet.  
Additional azimuthal features that may exist when a massive planet interacts with the disk, such as vortices or 
spiral arms \citep[e.g.][]{kley2006, ataiee2013, fung2014, zhu2014, attila2014}, are not considered in this work. 
Instead, we focus on the radial distribution of particles. 

\subsection{Dust evolution and radial gas velocity} \label{dust_evo}

To model the dust evolution, we use the formulation of \cite{birnstiel2010}. 
This model solves the advection-diffusion differential equation for the dust surface density $\Sigma_d$, 
and simultaneously simulates the growth, fragmentation, and erosion of dust grains by considering collisions of particles.  
For the relative velocities between particles, Brownian motion, turbulent velocities, settling to the midplane, and radial 
dust velocities are taken into account. 
The detailed explanation of this dust evolution model is in \cite{birnstiel2010}.
This model has been used extensively to investigate dust distributions in different types of disks, 
including comparisons with observations \citep[e.g.][]{birnstiel2013, marel2013, pinilla2013, dejuanovelar2013}. 
We investigate the evolution of 180 species of dust grains, defined by size from 1~$\mu$m to 200~cm. 
The radial velocity of dust depends on the coupling of the dust particles to the gas. 
The Stokes number (St) quantifies this coupling, which in the Epstein regime, where the mean free path of the 
gas molecules $\lambda_{\rm{mfp}}$ is higher than the size of particles $a$ ($\lambda_{\rm{mfp}}\geq 4/9 a$), 
is defined as

\begin{equation}
\rm{St}=\frac{a~\rho_s}{\Sigma}\frac{\pi}{2},
\label{stokes_number}
\end{equation}

\noindent where $\rho_s$ is the volume density of a dust grain of size $a$. 
In terms of the Stokes number, the total radial dust velocity is

\begin{equation}
	\varv_{\mathrm{r,d}}=\frac{\varv_{\mathrm{r,g}}}{1+\textrm{St}^2}+\frac{1}{\textrm{St}^{-1}+\textrm{St}} \frac{\partial_r P}{\rho_g \Omega},
 \label{dustvel} 
\end{equation}

\noindent with  $\rho_g$ being the total gas surface density. 
The first term of Eq.~(\ref{dustvel}) depends on the radial gas velocity $\varv_{\mathrm{r,g}}$, 
which in a disk evolving by viscous accretion is

\begin{equation}
	\varv_{\mathrm{r,g}}=-\frac{3}{\Sigma \sqrt{r}}\frac{\partial}{\partial r}(\Sigma\nu\sqrt{r}).
  \label{gas_radial_vel}
\end{equation}

\begin{figure*}
 \centering
   \includegraphics[width=18.5cm]{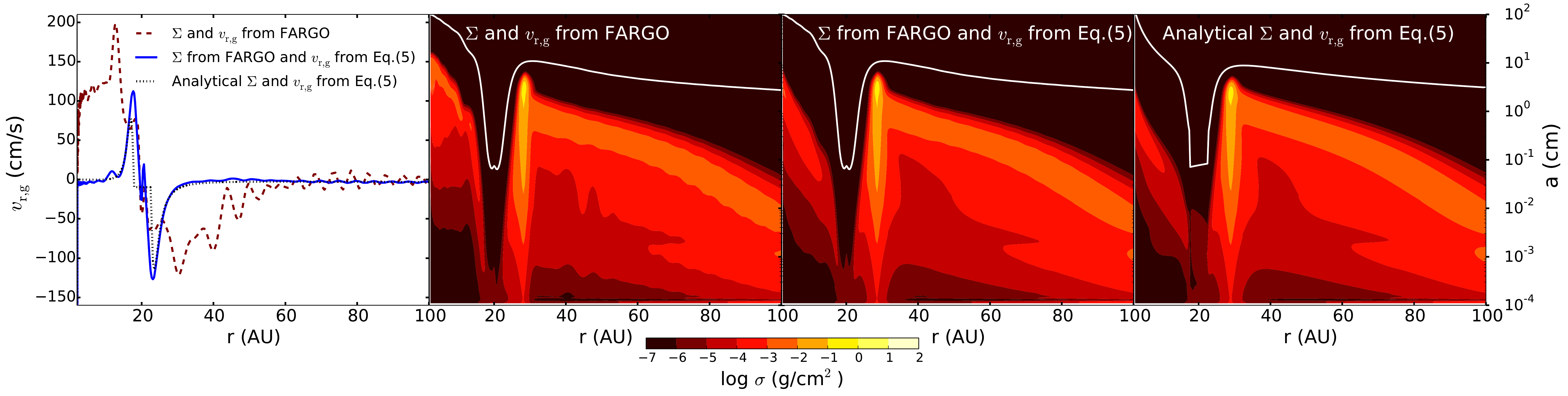}
   \caption{\emph{Left panel:} comparison of the gas radial velocity obtained by hydrodynamical simulations using \texttt{FARGO} for $q=1\times10^{-3}$,  $H/r=h_0r^f$ with $h_0=0.05$ and $f=0.25$, and $\alpha=10^{-3}$, when $\varv_{\mathrm{r,d}}$ is calculated from viscous accretion (Eq.~\ref{gas_radial_vel}), assuming $\Sigma$ by azimuthally averaging the gas surface density from \texttt{FARGO} simulations, and when the gas surface density is calculated analytically (Sect.~\ref{carved_gaps}) and $\varv_{\mathrm{r,d}}$  from Eq.~(\ref{gas_radial_vel}). \emph{Other panels:} dust density distribution at the same time of evolution ($\sim~1$Myr) for the same assumptions. White line corresponds to $\rm{St}=1$ (Eq.~\ref{stokes_number}), which is proportional to the gas surface density.}
   \label{comparison_velocity}
\end{figure*}

When the dust evolution models are combined with hydrodynamical simulations, the gas surface density and gas radial velocities can be directly taken as initial conditions for the dust evolution models, once the disk has reached a steady-state. Alternatively, the gas surface density can be based on the hydrodynamical simulations and the gas radial velocity from Eq.~(\ref{gas_radial_vel}) as in \cite{pinilla2012}. When a non-migrating planet is in the disk, $\Sigma$ is calculated with the analytical solution described in Sect.~\ref{carved_gaps} and using this $\Sigma$ -profile, the gas radial velocities are calculated assuming Eq.~(\ref{gas_radial_vel}). To see the potential differences of assuming the gas radial velocities from hydrodynamical simulations or viscous accretion, Fig.~\ref{comparison_velocity} compares the gas radial velocities obtained by hydrodynamical simulations using \texttt{FARGO} for $q=1\times10^{-3}$,  $H/r=h_0r^f$ with $h_0=0.05$ and $f=0.25$, and $\alpha=10^{-3}$ (azimuthally averaged and averaging over the last 100 orbits of evolution i.e orbits 1000-1100), when $\varv_{\mathrm{r,d}}$ is calculated from viscous accretion (Eq.~\ref{gas_radial_vel}) and assuming $\Sigma$ from \texttt{FARGO}; and when the gas surface density is calculated analytically (Sect.~\ref{carved_gaps}) and $\varv_{\mathrm{r,d}}$  from Eq.~(\ref{gas_radial_vel}). In addition the dust density distributions after $\sim~1$~Myr of evolution are displayed for each case. The main differences of the gas radial velocities are close to the planet where strong perturbation waves due to the planet create a wiggle profile. However, the dust density distribution is similar for each case, because radial drift and diffusion are more important than the drag term in Eq.~(\ref{dustvel}).  Hence, we have also proved the reliability of this analytical approach compared to the hydrodynamical simulations to obtain proper radial density distribution of dust particles.

\subsection{Visibilities at mm-wavelenghts}

To compare the results from the dust evolution models with millimetre observations, we calculate the real part of the 
visibilities in the $uv$-plane: $V_{\rm{Real}} (r_{uv})$ is given by \citep{berger2007}

\begin{equation}
	V_{\rm{Real}} (r_{\rm{uv}})=2\pi\int_0^{\infty} I(r) J_0(2\pi r_{{uv}}) r dr,
  \label{real_visi}
\end{equation}

\noindent where $J_0$ is the zeroth-order Bessel function of the first kind and $I(r)$ is the 
radial-dependent emergent intensity, which is directly calculated using the 
vertically integrated dust density distribution $\sigma (r,a)$ from the dust evolution results at a specific time of evolution. 
Thus, $\sigma (r,a)$ is different at each location for all the dust particle sizes (180 species) assumed in this work. 
For a given wavelength ($\lambda$), the intensity is given by,

\begin{equation}
	I(r)=B_\lambda (T (r))\left[1-\exp^{-\tau_\lambda(r)}\right],
  \label{intensity}
\end{equation}

\noindent with $B_\lambda (T (r))$ being the Planck function at the temperature $T(r)$ and $\tau_\lambda$ the optical depth, which is computed as

\begin{equation}
	\tau_\lambda=\frac{\sigma (r,a)\kappa_\lambda}{\cos i},
  \label{optical_depth}
\end{equation}

\noindent where the opacities at a particular wavelength $\kappa_\lambda$  are calculated for each grain size, 
assuming Mie theory and a mix of magnesium-iron silicates \citep[e.g.][]{dorschner1995}. 
The optical constants are taken from the Jena database \footnote{http://www.astro.uni- jena.de/Laboratory/Database/databases.html, 
with a specific silicate composition of: 10\% MgFeSiO$_4$, 28\% MgSiO$_3$, 31\% Mg$_2$SiO$_4$, 1\% NaAlSi$_2$O$_6$.}.
 
\subsection{Set-up} \label{setup}

\begin{table}
\caption{Stellar, disk and planets parameters}
\centering   
\tabcolsep=0.08cm                      
\begin{tabular}{c|c|c}       
Parameter &Symbol /units&  Value \\
\hline
\hline
Stellar mass&$M_{\star}[M_{\odot}]$& $2.4$\\
Disk mass&$M_{\rm{disk}}[M_{\odot}]$& $0.05$\\
Inner disk radius&$r_{\rm in}[\rm{AU}]$ & $2.0$\\
Outer disk radius&$r_{\rm out}[\rm{AU}]$& $400$\\
Viscosity&$\alpha$& $[2\times10^{-3}, 5\times10^{-3}]$\\
Inner planet mass&$M_{p1}[M_{\rm{Jup}}]$&$[1, 5, 10, 20, 30]$\\
Outer planet mass&$M_{p2}[M_{\rm{Jup}}]$&$[5, 10, 15, 20]$\\
Inner planet position&$r_{p1}[\rm{AU}]$& $\sim10$\\
Outer planet position&$r_{p2}[\rm{AU}]$& $\sim70$\\
Distance to the disk&$d$[pc]&103\\
Disk inclination&$i[^\circ]$ &45\\
Fragmentation velocity&$\varv_f[\rm{m~s}^{-1}]$ &10\\
Volume density of dust&$\rho_s[ \rm{g~cm}^{-3}]$&1.2\\
\hline
\hline
\end{tabular}    
\label{parameters}
\end{table}

We assume the disk mass to be $0.05~M_{\odot}$, consistent with the dust mass from \cite{mulders2013}, 
and a canonical dust-to-gas mass ratio of 100; however, disk masses are very uncertain. 
From optically thin mm emission, the disk mass can be estimated assuming dust opacities and a dust-to-gas mass ratio \citep[e.g.][]{andrews2005}. 
Nonetheless, this calculation is unreliable since the dust opacity depends on composition and shape of the dust \citep[e.g.][]{min2005, demyk2013}, 
and the gas and dust are not necessarily co-spatial \citep[e.g.][]{birnstiel2014}. 
The gas mass can also be calculated from observations of CO and its isotopologues \citep[e.g.][]{williams2014}; 
however, this estimation can depend on the chemical disk evolution, such as isotope selective processes \citep{miotello2014}. 
The disk mass assumed for HD~100546 is broadly consistent with the mass estimate from e.g. \cite{henning1998}, 
but it remains a very uncertain parameter and thus we keep it fixed in this work.

\begin{figure}
 \centering 
   	\includegraphics[width=8.5cm]{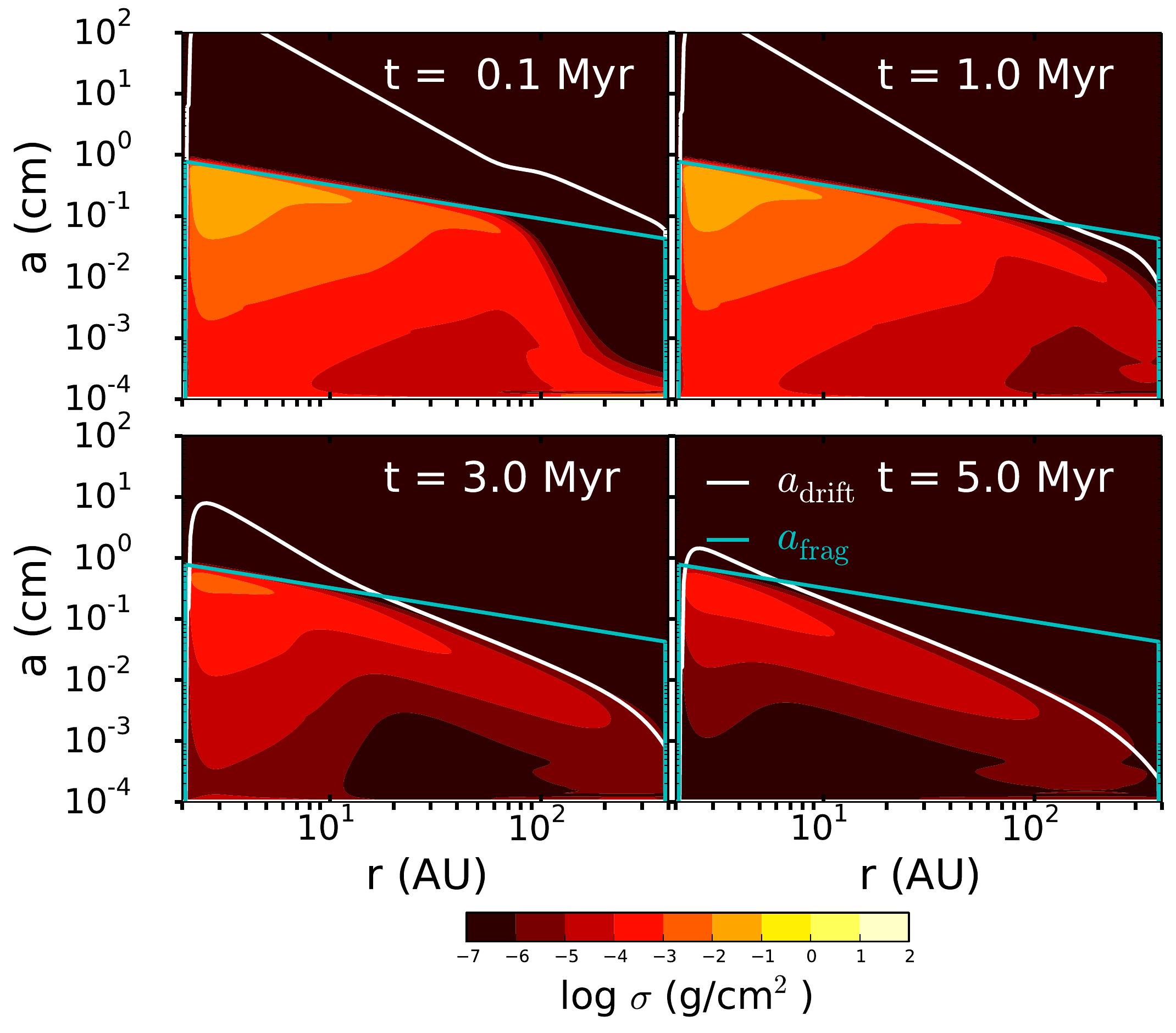}\\
	\includegraphics[width=8.5cm]{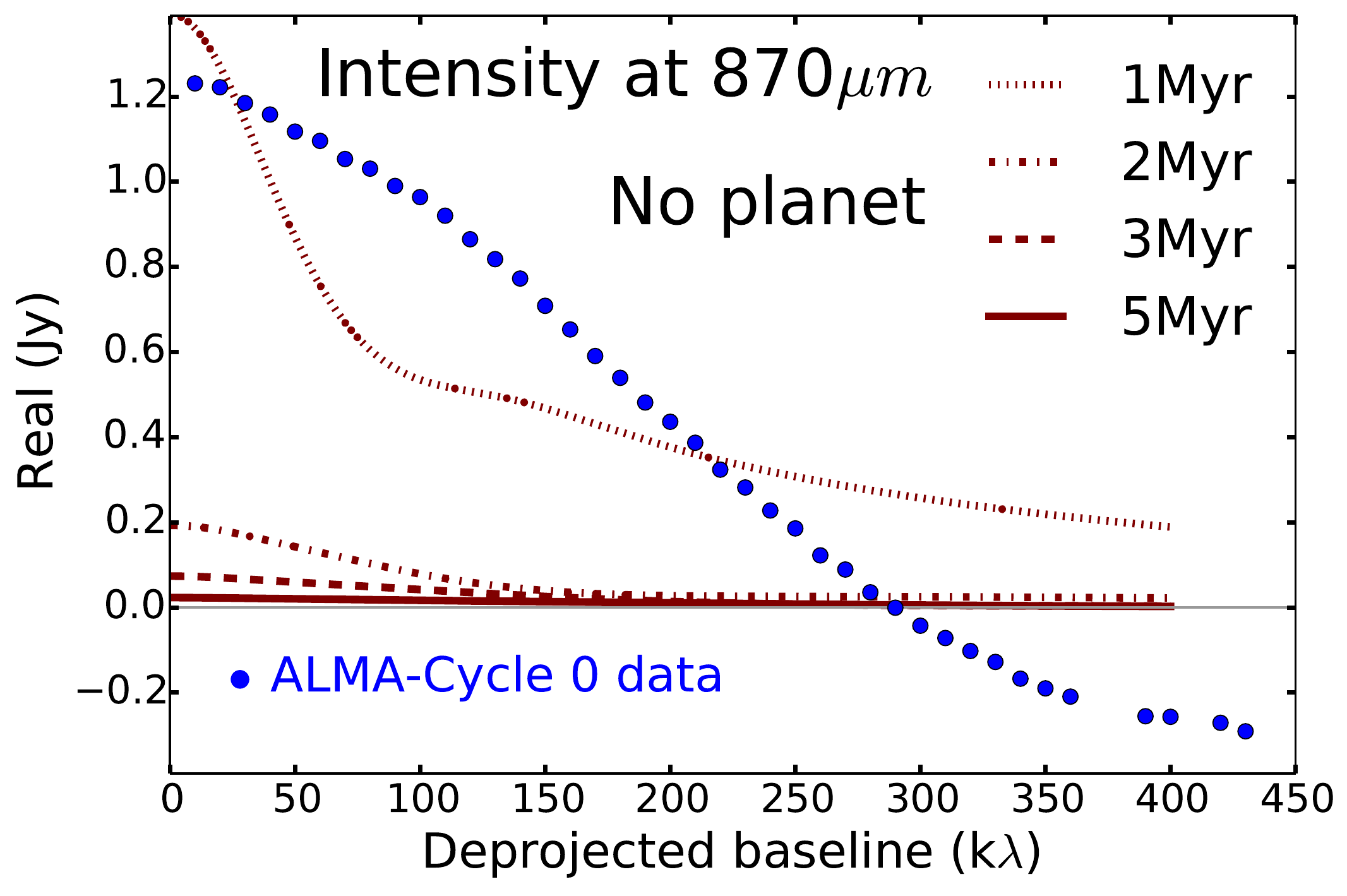}
   \caption{Dust density distribution at different times of evolution (\emph{top panel}) and real part of the visibilities at $870\mu$m (\emph{bottom panel}) at different times of dust evolution, when no planet is embedded in the disk. ALMA Cycle~0 data at this wavelength are over-plotted \citep[data from][]{walsh2014}. In the upper panel, the white and blue line correspond to $a_{\rm{drift}}$ (Eq.~\ref{adrift}), and $a_{\rm{frag}}$ (Eq.\ref{afrag}), respectively.}
   \label{no_planet_plot}
\end{figure}

We fix the gas disk extent to $[2-400]$~AU, in agreement with the CO J = 3-2 emission from \cite{walsh2014}. 
The initial gas surface density is a power law, such that $\Sigma(r)=\Sigma_0 (r/r_{p1})^{-1}$.  
We use a temperature profile specific for HD~100546 and derived by \cite{bruderer2012}, 
who constrained the gas temperature via detailed modelling of the observed low-J CO line emission from single dish observations with the Atacama Pathfinder Experiment (APEX) \citep{panic2010}, mid/high-J CO lines observed with PACS-Herschel \citep{sturm2010},  and continuum emission. For the dust evolution models, we impose the dust temperature as the gas temperature in the midplane, where $T_{\rm{gas}}\simeq T_{\rm{dust}}$ (with values of $T_{20\rm{AU}}\simeq 100$K, $T_{400\rm{AU}}\simeq 25$K). This temperature is an upper limit, since it is obtained from observations of the warm gas in the disk atmosphere. However, CO channel maps from ALMA observations \citep{walsh2014} do not show the double lobe signature, which indicates the presence of a CO freeze-out zone \citep[see e.g.][]{gregorio2013}. Hence, 
HD~100546 is a warm disk ($T_{\rm{midplane}}\gtrsim 20-25$K everywhere). Thus, we do not expect that the disk temperature considerably differs from the adopted temperature. 
This temperature is used for the dust evolution models and assumed to be the temperature of the vertically integrated dust density distribution  for the calculation of the intensity and visibilities at mm-wavelengths, which are mostly sensitive to the distribution of large grains. Because large (mm-sized) grains efficiently settle towards the midplane,  the surface density of large grains in the disk atmosphere (where the temperature is higher) is negligible \citep{dullemond2005}.  Therefore, adopting the midplane temperature only in the calculation of the emergent continuum emission is an appropriate assumption.

Based on that temperature, the aspect ratio is calculated as $c_s/v_K=H/r=h_0r^{f}$, obtaining $h_0=0.045$ and $f=0.33$ 
at the position where the inner planet is located (10~AU). 
This aspect ratio is directly used for obtaining the analytical shape of carved gaps. 
The disk viscosity is assumed to be  $\nu=\alpha c_s^2/\Omega$, with $\alpha=[2\times10^{-3}, 5\times10^{-3}]$ as in \cite{mulders2013}. 
For the visibilities, the disk is taken to be at a distance of 103~pc  with an inclination ($i$) of $45^\circ$ \citep{vandenAncker1997}, 
and a position angle (P.A.) of  $146^\circ$ (east from north). 

We assume a large range of planet masses according to previous studies \citep{acke2006, tatulli2011, mulders2013, quanz2013, currie2014}. 
For an inner companion located $r_{p1}\sim10$~AU,  we assume the planet-stellar mass ratio to be 
$\sim[4.2\times10^{-4}, 2.1\times10^{-3}, 4.2\times10^{-3}, 8.3\times10^{-3}, 1.3\times10^{-2}]$, 
which corresponds to $M_{p1}=[1, 5, 10, 20, 30]~M_{\rm{Jup}}$ planets around a $2.4~M_\odot$ star. 
For the outer planet located at $r_{p2}\sim70$~AU, we consider $M_{p2}=[5, 10, 15, 20]~M_{\rm{Jup}}$.

Finally, for the dust evolution models, we assume that the velocity threshold above which particles fragment, 
the so-called fragmentation velocity, is $\varv_f=10~\rm{m~s}^{-1}$. 
For the initial dust density distribution, we assume all particles to be $1~\mu$m-sized. 
The dust particles are considered to have a volume density of $\rho_s=1.2 \rm{g~cm}^{-3}$, 
according to the averaged values of the volume density for silicates \citep[e.g.][]{blum2008}. 
All parameters are summarised in Table~\ref{parameters}.

\section{Results}     \label{results}

In this section, we present the results for the case of no planet, and for the cases where either one or two planets are embedded in the disk. 
For the latter case, we consider the two scenarios where the second (outer) planet is injected either at the same time as the inner planet, 
or at a later time. 

For the models where the planets are injected at later times, we do not assume an actual model to introduce the planet and let it grow. 
Our main aim is to study the final dust density distributions assuming that the planets are already formed and embedded in the disk. 
Massive planets such as those assumed in this work are expected to have slow type II migration, in which case migration timescales 
follow the viscous diffusion time of the gas, as do the pressure bumps and the dust. 
Thus, potential dust traps are likely to follow the migration of the planets, keeping qualitatively similar results. 

\subsection{No planets in the disk} \label{no_planet}

Figure~\ref{no_planet_plot} illustrates the dust density distribution at different times of evolution, 
and the corresponding real part of the visibilities at $870\mu$m, when no planet is embedded in the disk. 
Moreover, ALMA Cycle~0 data at this wavelength are also plotted for comparison. 
Note that the errors are also plotted and they are smaller than the point size 
\citep[because of the very high signal-to-noise ratio of the ALMA data, ][]{walsh2014}. 
For this simulation $\alpha=2\times10^{-3}$ is assumed.  
At early times ($t\lesssim~1$~Myr), the dust particles grow to the maximum grain size before particles fragment. 
When fragmentation is mainly because of turbulent relative velocities, $a_{\rm{frag}}$ is \citep{birnstiel2012},

\begin{equation}
	a_{\rm{frag}}=\frac{2}{3\pi}\frac{\Sigma}{\rho_s \alpha}\frac{\varv_f^2}{c_s^2}.
  \label{afrag}
\end{equation}

At later times, the dust surface density decreases, so the drift barrier moves to smaller sizes and possibly 
below the fragmentation size (Eq.~\ref{afrag}). In such cases, $a_{\rm{drift}}$ is \citep{birnstiel2012}

\begin{equation}
	a_{\rm{drift}}=\frac{2 \Sigma_d}{\pi\rho_s}\frac{v_K^2}{c_s^2}\left \vert \frac{d \ln P}{d\ln r} \right \vert^{-1}.
  \label{adrift}
\end{equation}

\begin{figure*}
 \centering
 \tabcolsep=0.03cm 
   \begin{tabular}{cc}   
   	\includegraphics[width=9cm]{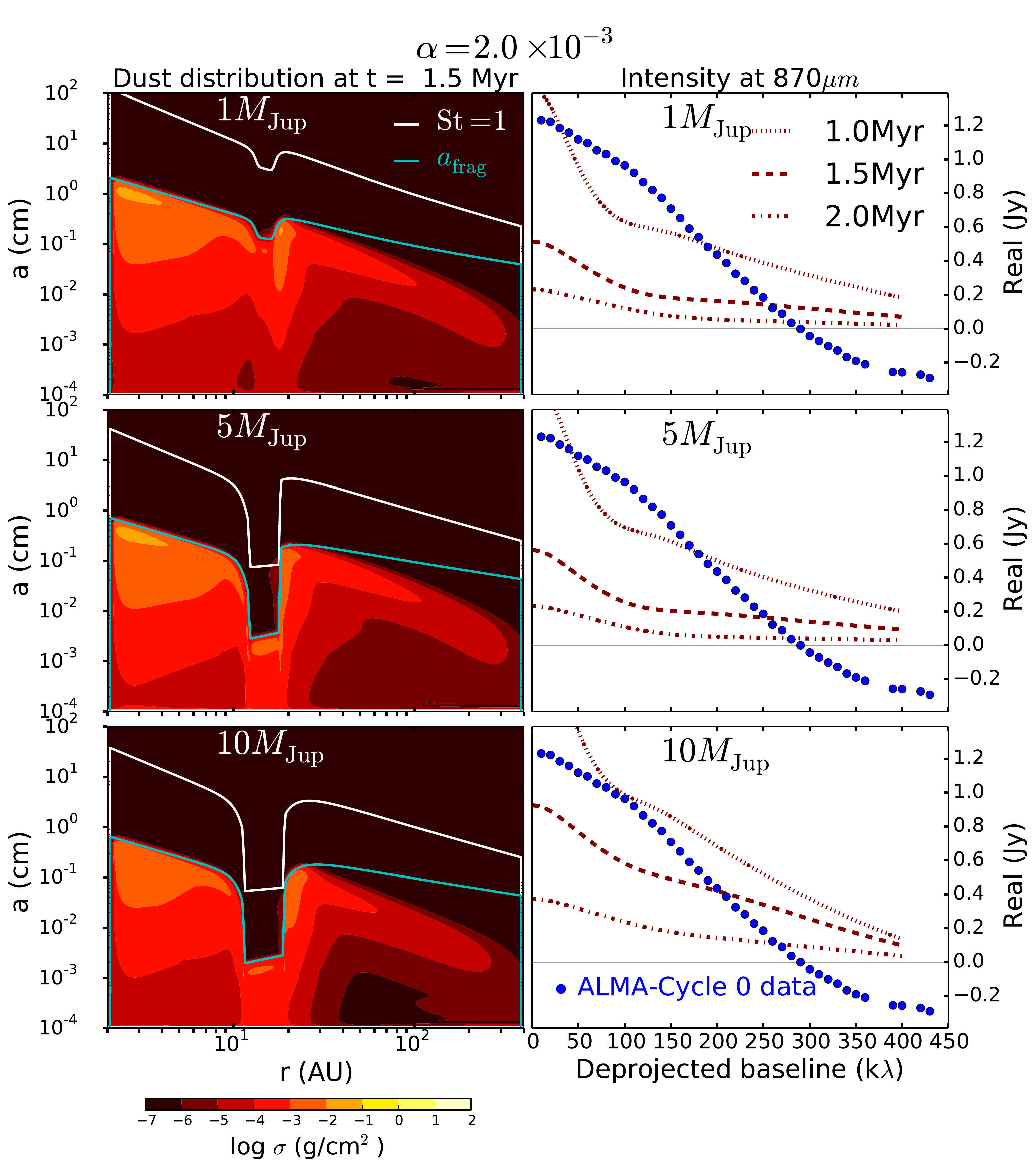}&
	\includegraphics[width=9cm]{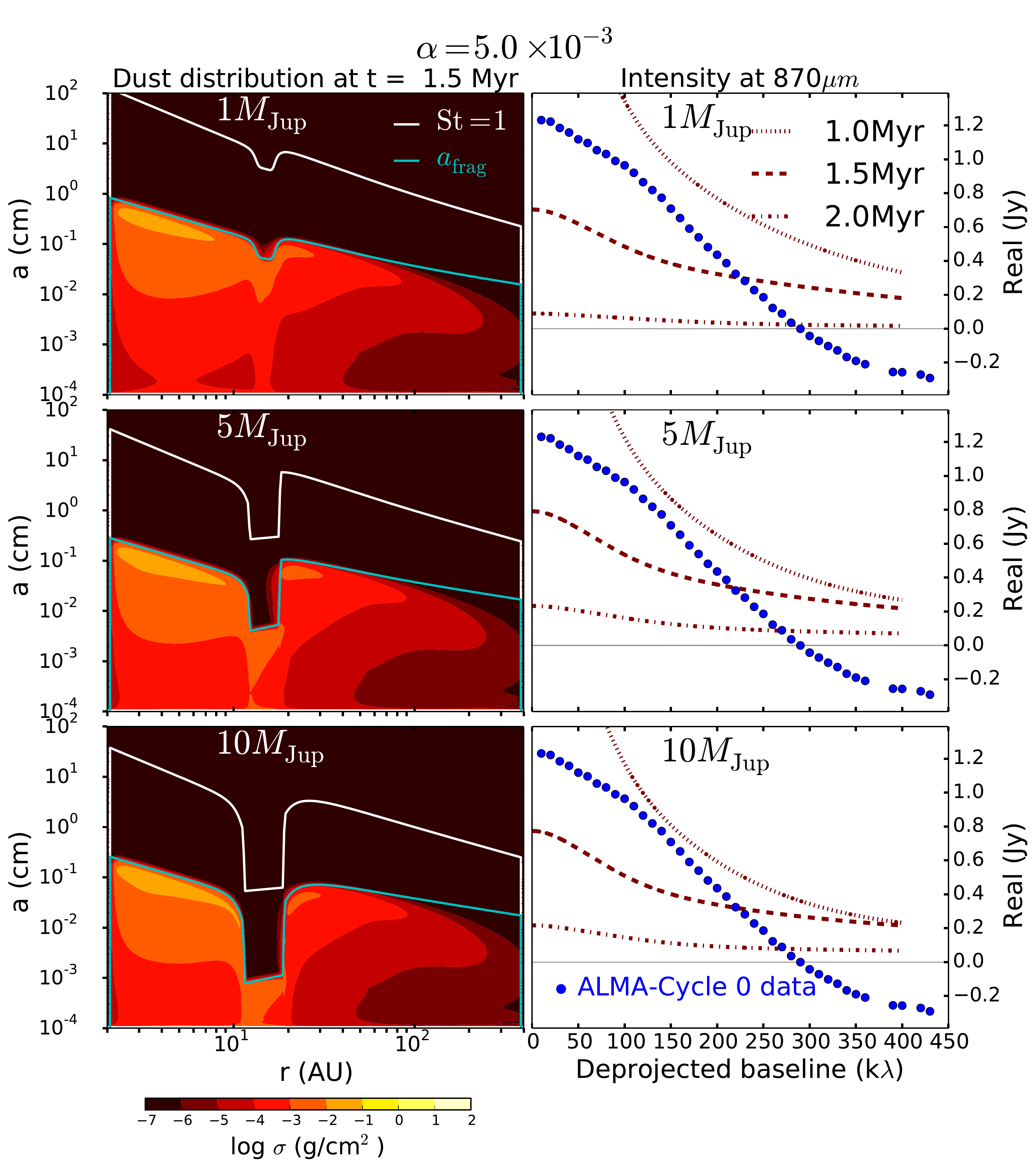}
   \end{tabular}
   \caption{Dust density distribution after 1.5~Myr of evolution  and real part of the visibilities at $870\mu$m  at different times of dust evolution, when a single planet interacts with the disk for two different values of disk viscosity, $\alpha=2\times10^{-3}$ ({\em left column})  and $\alpha=5\times10^{-3}$ ({\em right column}). From top to bottom, the masses of the planet are: 1, 5, and $10~\rm{M_{\rm{Jup}}}$. ALMA Cycle~0 data are over-plotted in Fig.~\ref{no_planet_plot}. In the dust density distribution plots, the white and blue lines correspond to $\rm{St}=1$ (Eq.~\ref{stokes_number}), and $a_{\rm{frag}}$ (Eq.~\ref{afrag}) respectively.}
   \label{one_planet_1_5_10Mjup}
\end{figure*}

Before particles start to drift, millimetre particles are distributed in the entire disk (Fig.~\ref{no_planet_plot}), 
creating a visibility profile in disagreement with observations, which suggest a more concentrated ring centred at $\sim26$AU 
($\sim290~\rm{k}\lambda$) \citep{walsh2014}. 
Once the particles grow to sizes for which radial drift dominates ($t\gtrsim~1$~Myr), 
the disk is quickly depleted of millimetre dust particles. 
This is reflected in the profile of the real part of the visibilities where the flux at $870\mu$m drastically reduces after 1~Myr of evolution, 
becoming already very low at $2$~Myr. 

As a consequence, any mechanism that helps to reduce the rapid inward drift is needed to explain the millimetre emission of this disk. 
We assume that a pressure trap is formed at the outer edge of a gap carved by a massive planet, motivated also by the companion 
candidates suggest in the literature \citep[e.g.][]{acke2006, tatulli2011, brittain2014}. 
Particles concentrate in the regions of highest pressure \citep[e.g.][]{Weidenschilling1977, Brauer2008}, 
and therefore the assumed pressure bump can help to reduce the radial drift. 

\subsection{A planet is embedded in the inner disk} \label{one_planet_section}
Assuming the analytical profiles for gaps carved by massive planets described in Sect.~\ref{carved_gaps}, 
we consider an inner planet of different masses $M_{p1}=[1, 5, 10, 20, 30]~M_{\rm{Jup}}$  at $r_{p1}\sim10$AU to 
investigate the trapping efficiency and the comparison with observations. 
Since we neglect the possible effect of an eccentric gap in our simulations,  
we give an approximated value for the location of the planets (Sect~\ref{carved_gaps}).

\begin{figure*}
 \centering
 \tabcolsep=0.001cm 
   \begin{tabular}{cc}   
   	\includegraphics[width=8.4cm]{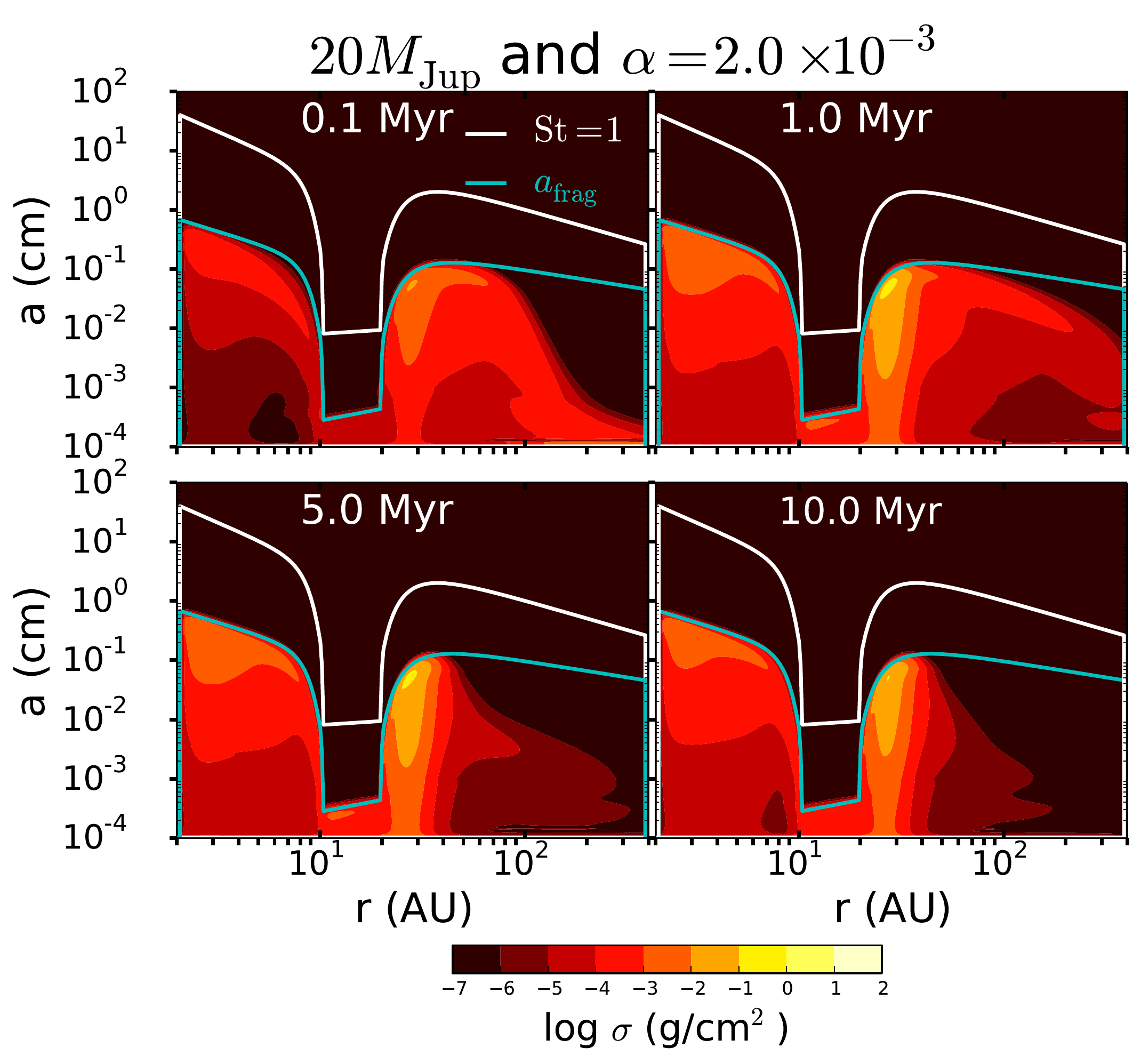}&
	\includegraphics[width=8.4cm]{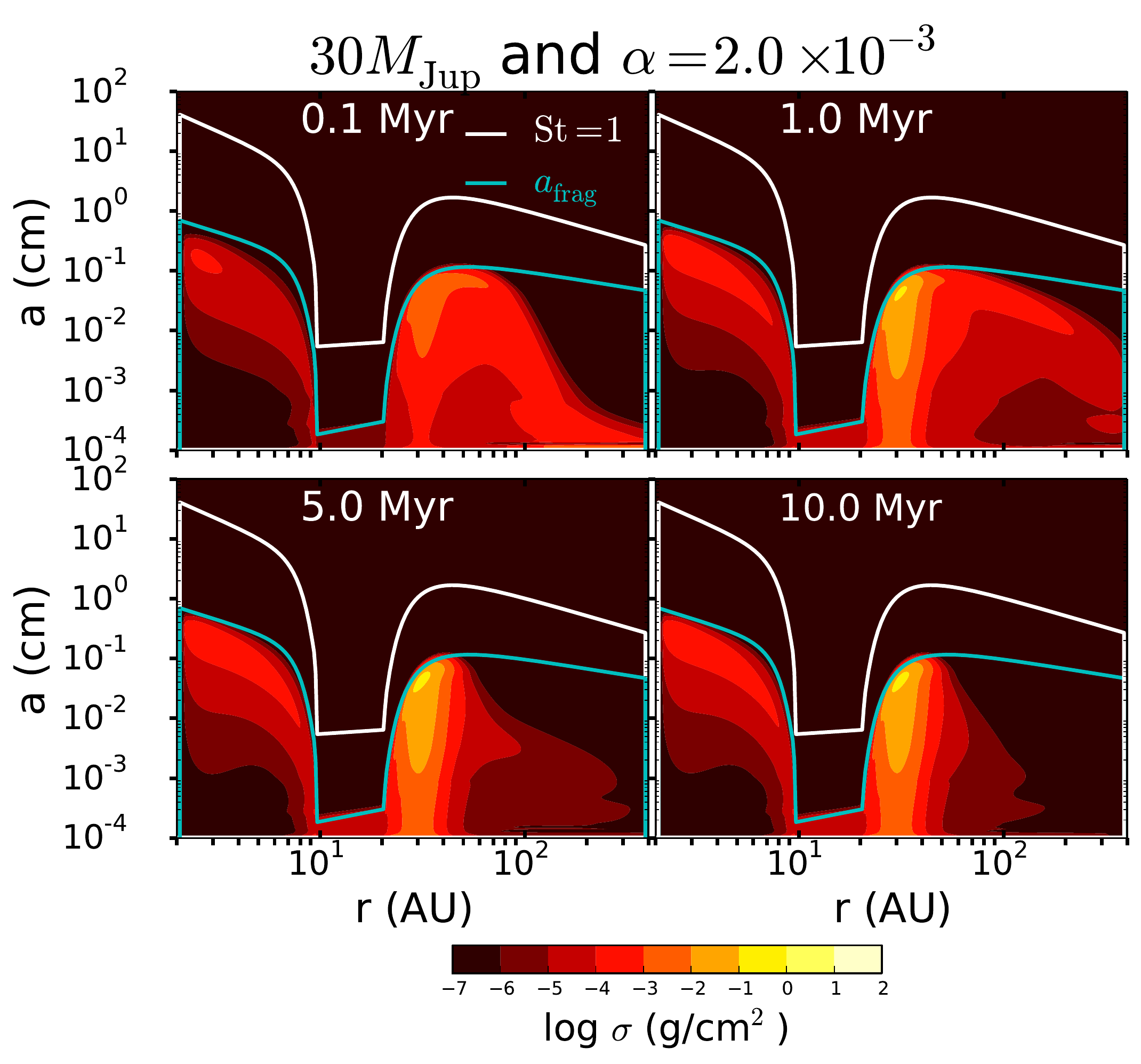}\\
	[-1.6ex]
	\includegraphics[width=9.0cm]{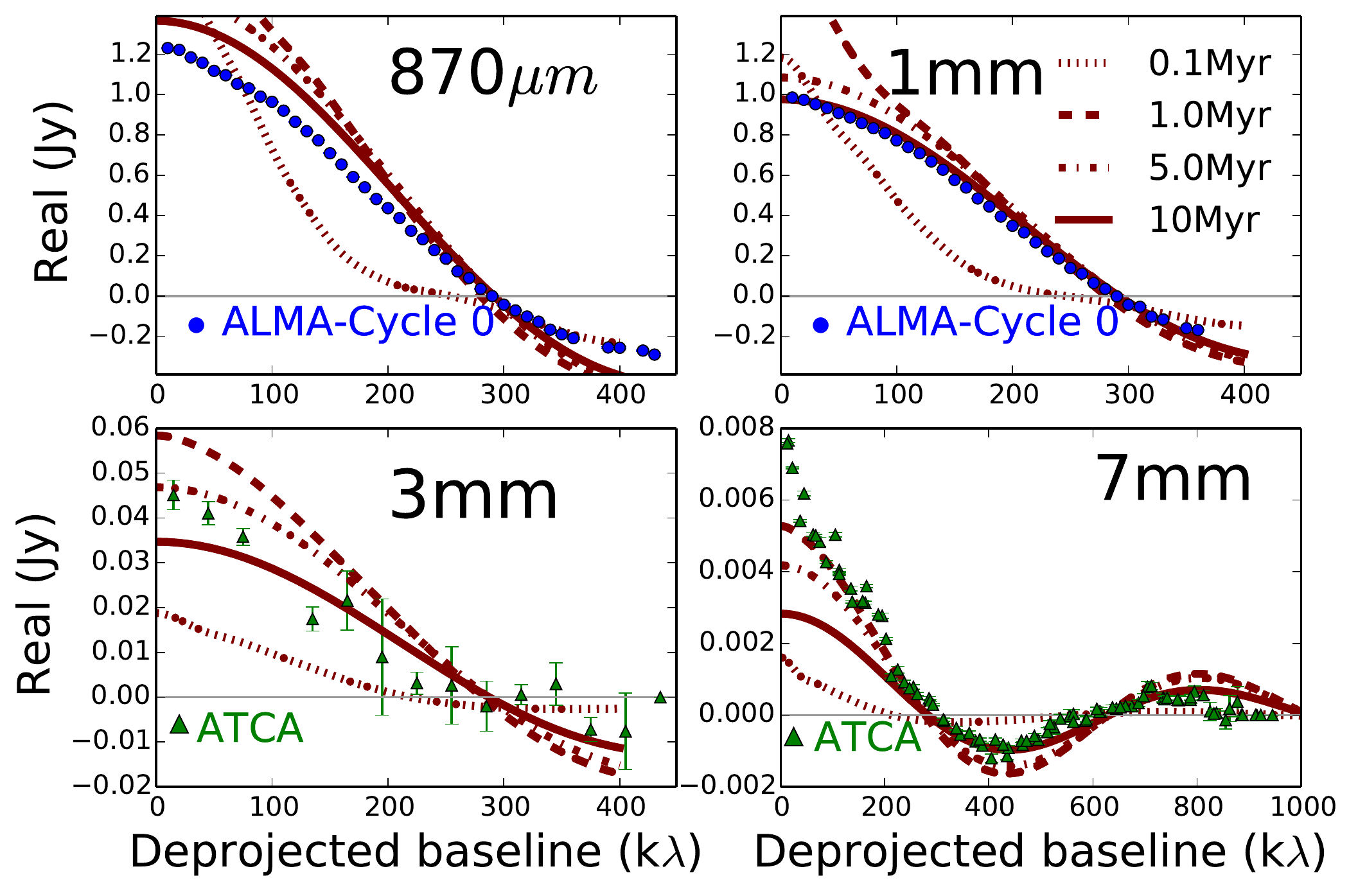}&
	\includegraphics[width=9.0cm]{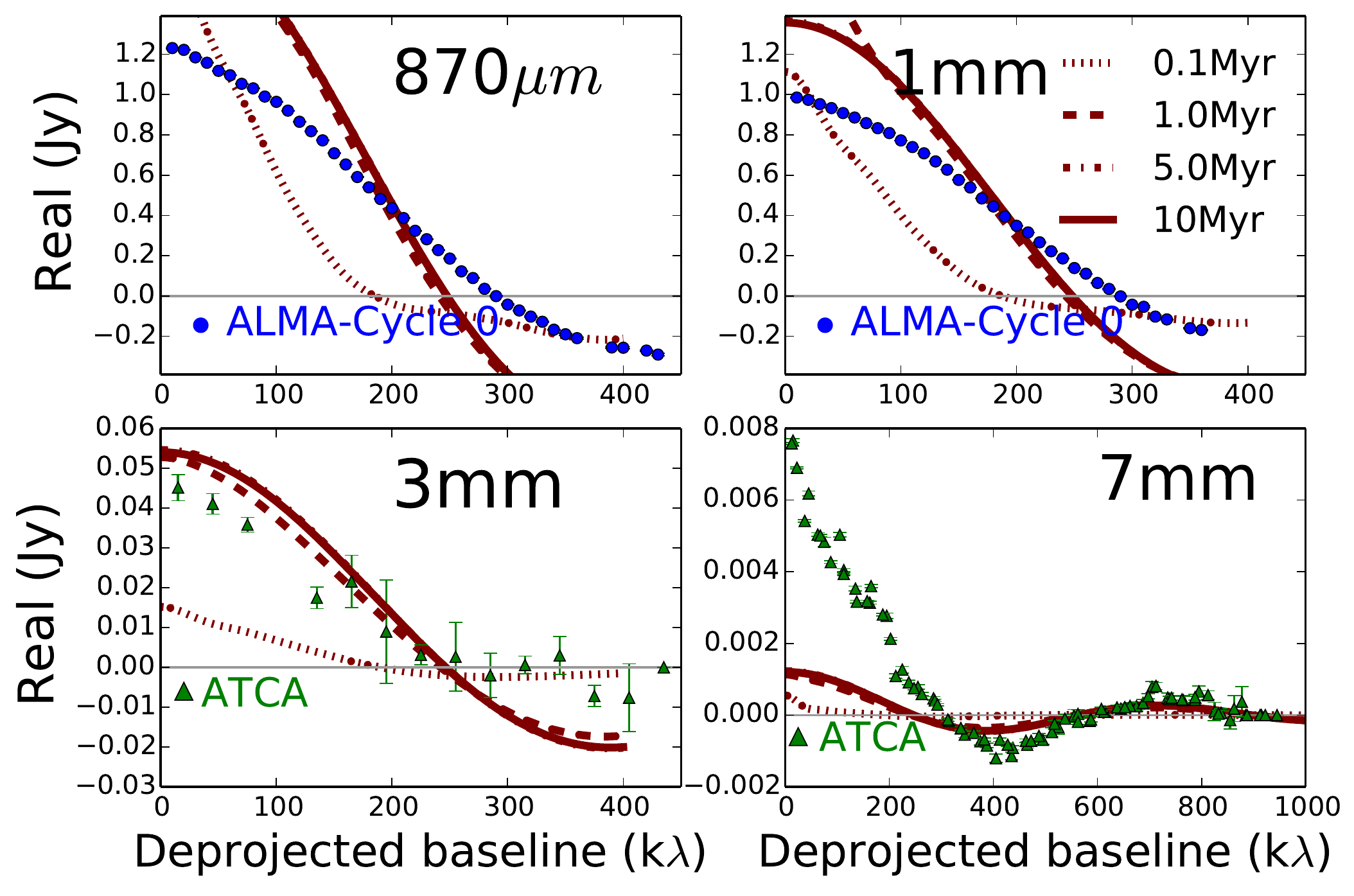}
   \end{tabular}
   \caption{Dust density distribution at different times of evolution ({\em upper row}) and the corresponding real part of the visibilities at $\lambda=[0.87 , 1.0, 3.0, 7.0]$~mm ({\em lower row}) at different times of dust evolution, when a  $20~\rm{M_{\rm{Jup}}}$ planet (\emph{left column}) and $30~\rm{M_{\rm{Jup}}}$ planet  (\emph{right column}) is embedded in the inner disk. Data from ALMA Cycle~0 from \cite{walsh2014} and ATCA from \cite{wright2015} are over-plotted. In the upper panels, the white and blue line correspond to $\rm{St}=1$ (Eq.~\ref{stokes_number}), and $a_{\rm{frag}}$ (Eq.\ref{afrag}) respectively.}
   \label{20_30Mjup_inner_planet_plot}
\end{figure*}

Figure~\ref{one_planet_1_5_10Mjup} shows the dust density distribution after 1~Myr of evolution, 
and the corresponding real part of the visibilities at $870\mu$m for $\alpha=[2\times10^{-3}, 5\times10^{-3}]$, 
when a 1, 5 or 10~$M_{\rm{Jup}}$ planet is embedded in the inner part of the disk. 
The results of these simulations suggest that a planet whose mass is lower than 10~$M_{\rm{Jup}}$ is not massive enough to result in trapping of millimetre-sized dust particles, sufficient to create a visibility curve that is in agreement with ALMA Cycle~0 observations. 
In any of these cases, the flux is already under-predicted after 1.5~Myr of evolution. 
\cite{pinilla2012} show that when a planet of $q=10^{-3}$, i.e. a  1~$M_{\rm{Jup}}$ around a solar-type star, 
carves a gap in the disk, particles can be trapped at the outer edge when $\alpha=10^{-3}$. 
The reason why trapping is more difficult in this case is because of the larger difference between the 
maximum grain size or $a_{\rm{frag}}$  (Eq.~\ref{afrag}), 
and the particle size in which dust feels the highest radial drift i.e. $\rm{St}=1$. 
Particles with $\rm{St}\sim1$ are the easiest to trap, however $a_{\rm{frag}}$ is in this case around one 
order of magnitude lower than the grain size that corresponds to $\rm{St}=1$. 
These smaller particles are more difficult to trap because the drift is inefficient for particles with 
$\rm{St}\lesssim \alpha$.  
For these models, the maximum grain size is lower because of the higher temperatures and 
$\alpha$-viscosity considered here compared with those in \cite{pinilla2012}. 
The visibility shape does not change significantly for the two values of the viscosity. 
For this reason, for the following results, we only focus on the cases with $\alpha=2\times10^{-3}$. 
Increasing $\alpha$ decreases the maximum grain size even further (Eq.~\ref{afrag}), 
making the particle trapping more difficult. 
Assuming a lower viscosity or a higher fragmentation velocity can help to trap the particles \citep{pinilla2014b}.

\begin{figure*}
 \centering
   \includegraphics[width=18.0cm]{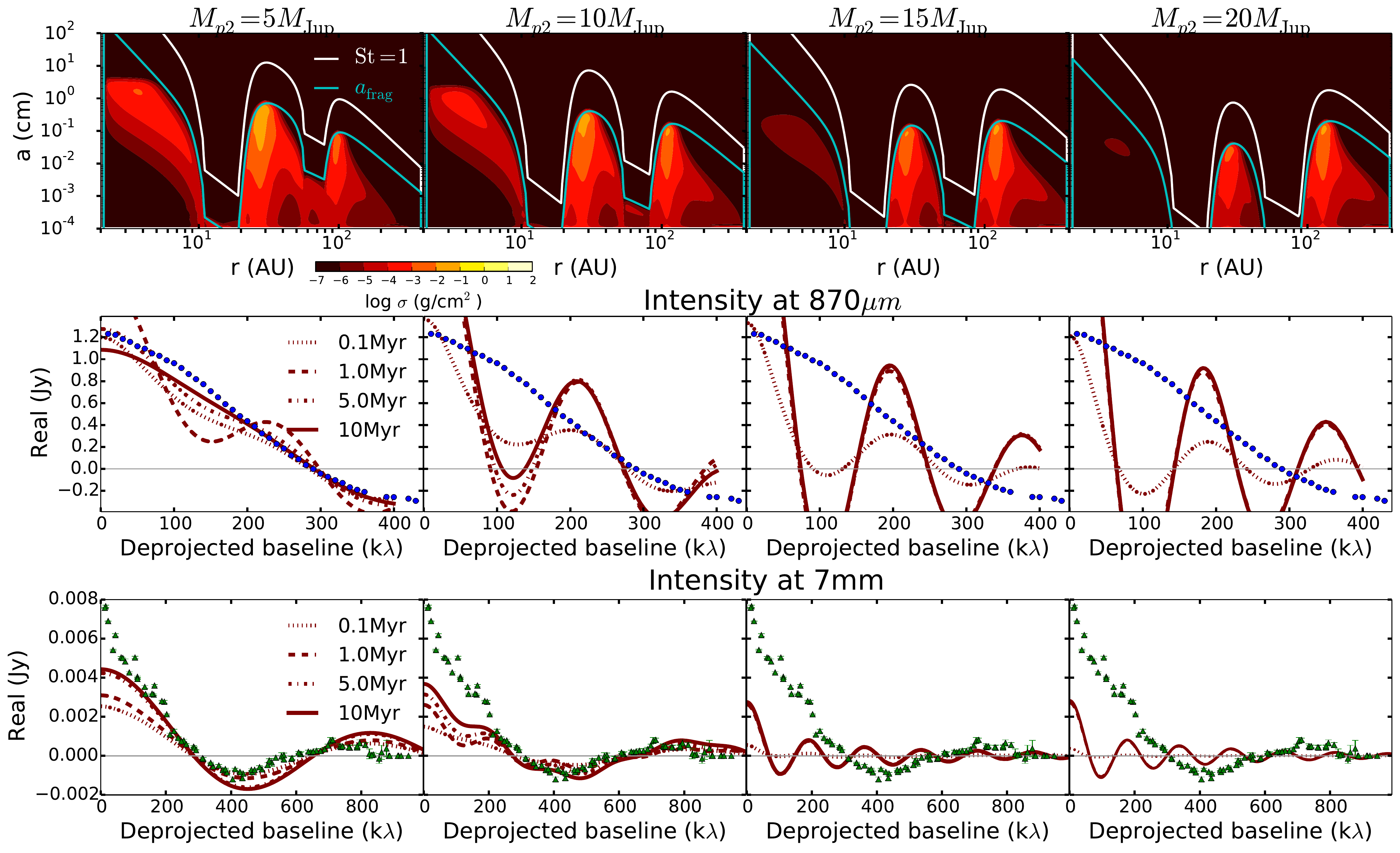}
     \caption{{\em Upper panels:} dust density distribution at 1.5~Myr of evolution when two planets are injected in the disk simultaneously. The mass of the inner planet is 20~$\rm{M_{\rm{Jup}}}$ and it is located at  $r_{p1}\sim10-15$AU, while the outer planet at $r_{p2}\gtrsim$70~AU is assumed to be $M_{p2}=[5, 10, 15, 20]~M_{\rm{Jup}}$ from left to right panels respectively. The white line is $\rm{St}=1$, and the blue line is $a_{\rm{frag}}$. The corresponding real part of the visibilities for each  $M_{p2}$ at $\lambda=0.87$mm ({\em middle panels}) and  $\lambda=7.0$mm ({\em lower panels}) are displayed for different times of dust evolution, for each $M_{p2}$. ALMA Cycle 0 and ATCA data are over-plotted for comparison.}
   \label{two_planets_same_time}
\end{figure*}

When the mass of the inner planet is increased, the trapping becomes more efficient. 
Figure~\ref{20_30Mjup_inner_planet_plot} shows the dust density distribution at different times of evolution and the 
real part of visibilities at $\lambda=[0.87 , 1.0, 3.0, 7.0]$mm, when a 20 and 30~$\rm{M_{\rm{Jup}}}$  planet is 
embedded in the inner part of the disk. 
ALMA Cycle~0 data  from \cite{walsh2014} and ATCA data from \cite{wright2015} are over-plotted for comparison. 
For the case of a 20~$\rm{M_{\rm{Jup}}}$ planet, the visibilities better fit the data; nonetheless, to obtain the 
best fit with the data, the dust needs to evolve for long timescales ($\gtrsim 5$~Myr) to grow, drift, and concentrate 
the millimetre grains in a narrow ring. 
Already at 5~Myr, most of the millimetre particles are concentrated in a narrow region; however, the large amount of these 
grains leads to a over-prediction of the fluxes. 
The continuous fragmentation that occurs in the pressure bump, together with some small-sized dust crossing the gap, 
reduce the amount of millimetre grains, enabling a better fit to the data at 10~Myr.  
The total flux is under-predicted at 7mm because the maximum grain size in the trap is around 1mm.  
The null of the visibilities at $\sim290~\rm{k}\lambda$ is in good agreement with the observations, 
implying that a 20~$\rm{M_{\rm{Jup}}}$ planet located in the inner part of the disk, can create a pressure maximum at $\sim$26AU, 
where the centre of the inner narrow ring was observed with ALMA Cycle~0.

If the planet is located at the same location, and its mass is increased to 30~$\rm{M_{\rm{Jup}}}$, the resulting carved gap is wider, 
moving the location of the pressure maximum outwards and therefore the peak of the millimetre emission. 
For this reason, the null of the visibilities is at shorter baselines ($\sim245~\rm{k}\lambda$, 
Fig.~\ref{20_30Mjup_inner_planet_plot}). 
In addition, the total flux is higher because the gap is also deeper, filtering more dust, and as a result, 
a larger number of millimetre particles remain in the trap. 
However, the fluxes at 7mm are lower than in the case of a 20~$\rm{M_{\rm{Jup}}}$ planet. 
This is because the maximum grain size has decreased slightly in the trap as the pressure maximum is further away, 
where the gas surface density is also lower, decreasing $a_{\rm{frag}}$  (Eq.~\ref{afrag}) 
within the bumps. 
An important remark is that at $7$~mm the emission likely has a $\gtrsim10\%$ contamination from free-free emission \citep{wright2015}, 
which is neglected in our calculations. Hence, we expect the models to underestimate the 7~mm flux.

From the simulations with one single planet in the inner disk, the best model that reproduces the observations at different wavelength 
is for a 20~$\rm{M_{\rm{Jup}}}$ planet and $\alpha=2\times10^{-3}$, 
consistent with the results from \cite{mulders2013}, who found similar results from modelling mid-infrared data.  
For these parameters, the best fit is for an old disk ($\sim5-10$~Myr). 
Nevertheless, with one single planet, the outer ring of mm emission observed with ALMA remains unexplained \citep{walsh2014} . 

\subsection{Two planets are embedded in the disk} \label{two_planets_section}
In this section, the results with two planets embedded in the disk are presented. 
We use the analytical shapes described in Sect.~\ref{carved_gaps} for both gaps. 
For all cases, the gas surface density is always scaled such that the disk mass is $0.05 M_{\odot}$. 
For the planet embedded in the inner disk, we consider the best single-planet fit of $M_{p1}=20~M_{\rm{Jup}}$, 
and a disk viscosity of $\alpha=2\times10^{-3}$. 
For the outer planet, it is assumed  $M_{p2}=[5, 10, 15, 20]~M_{\rm{Jup}}$ and $r_{p2}\sim70$AU. 
There are two sets of simulations: one where the two planets are assumed to interact with the disk from the same stage of evolution, 
and a second, where the outer planet is assumed to interact with the disk from a later stage.

\subsubsection{Simultaneous injection of both planets} \label{2planets_same_time}

Figure~\ref{two_planets_same_time} illustrates the dust density distribution at 1.5~Myr of evolution when two planets are interacting 
in the disk from the same stage of evolution ($t\sim1000$~yr). 
The inner planet is a 20~$\rm{M_{\rm{Jup}}}$ planet at  $r_{p1}\sim10$~AU, while the outer planet at 
$r_{p2}\sim$70~AU is assumed to be $M_{p2}=[5, 10, 15, 20]~M_{\rm{Jup}}$. 
In addition, the real part of the visibilities at $\lambda=[0.87 , 7.0]$mm are displayed for each $M_{p2}$, 
and at different times of dust evolution. 
When a second planet is assumed, the gas surface density sharply decreases with radius in the region between the two planets. 
As a consequence of the high negative pressure gradient, the dust particles move quickly inward (timescales shorter than $\sim$~1~Myr) 
and stop their migration in the pressure maximum at the outer edge on the inner gap. 
This effect creates a narrow ring-like concentration of large particles centred at $\sim~26$AU in shorter timescales 
than in the case of a single inner planet. 

\begin{figure*}
 \centering
     \tabcolsep=0.04cm 
   \begin{tabular}{cc}   
  	\includegraphics[width=8.0cm]{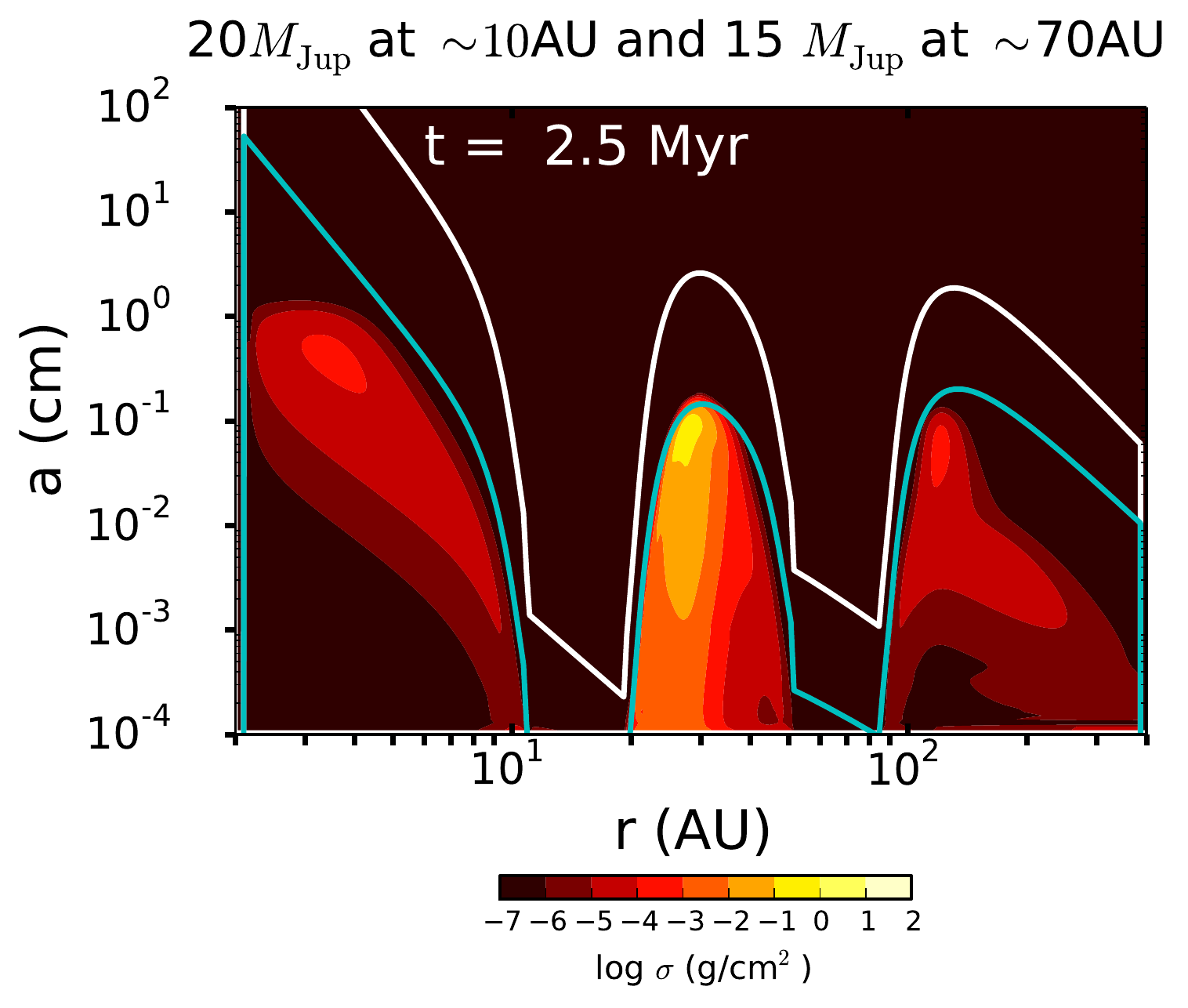}&\includegraphics[width=8.0cm]{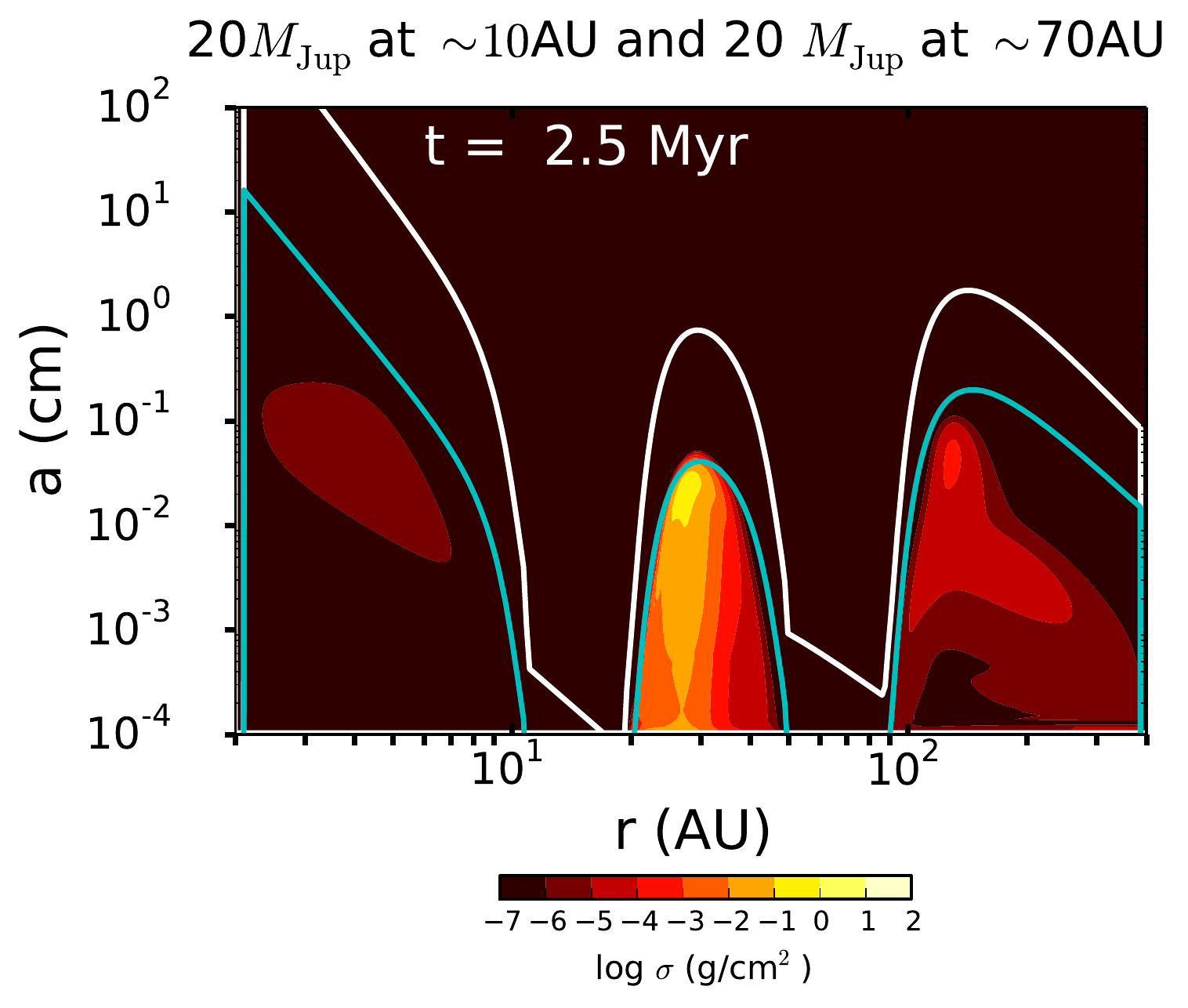}\\[-1.6ex]
	\includegraphics[width=8.0cm]{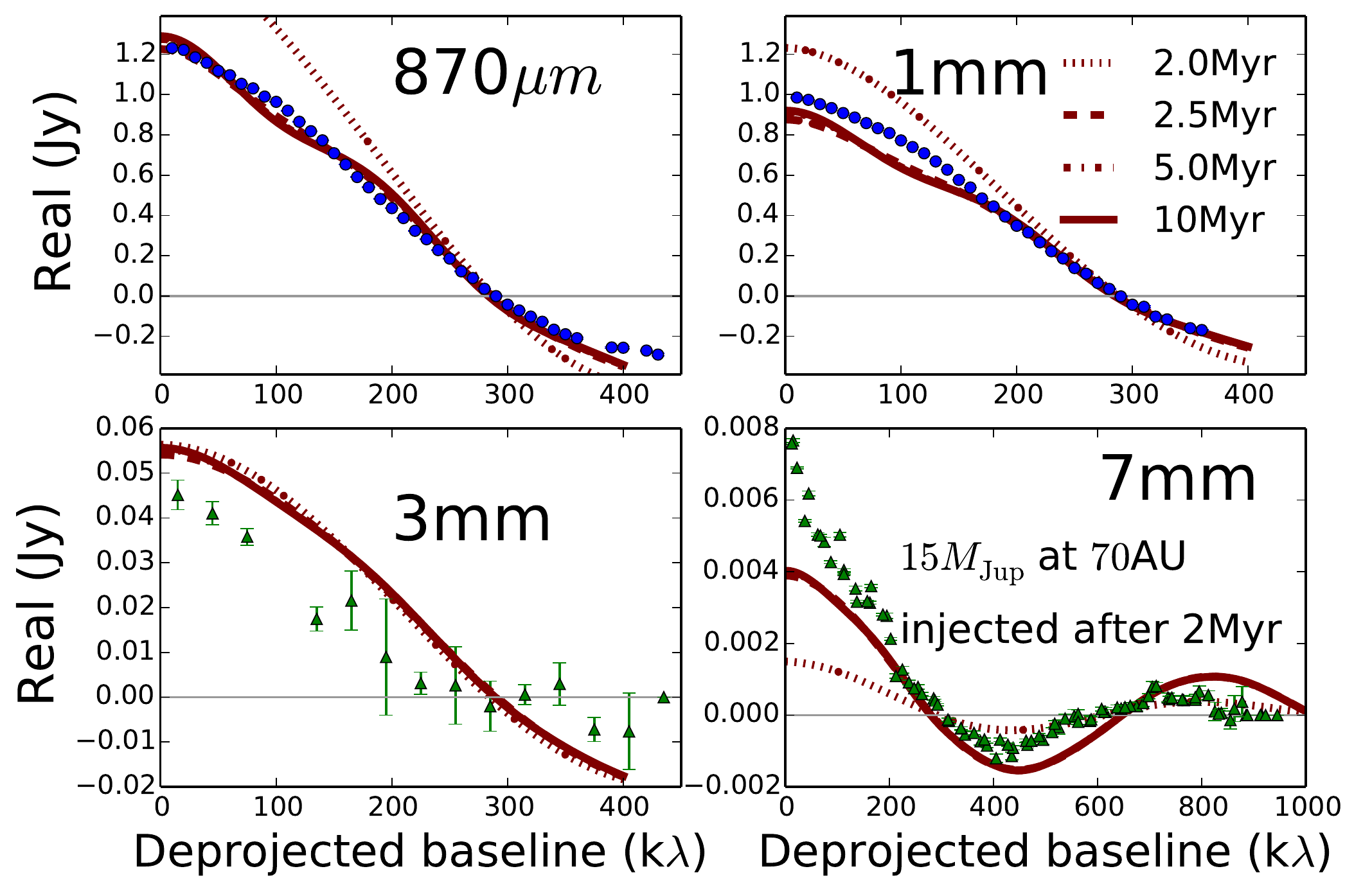}&\includegraphics[width=8.0cm]{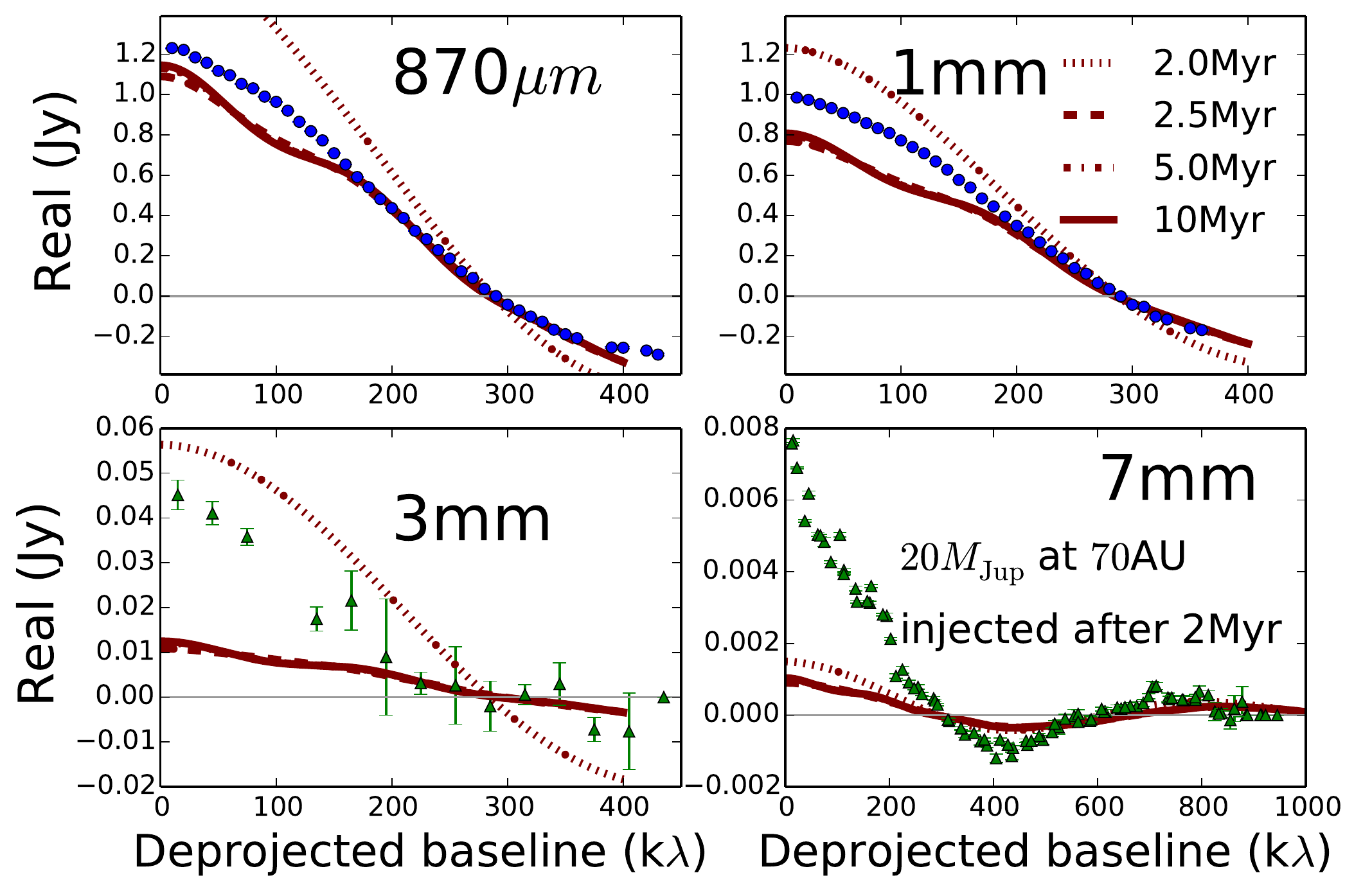}\\
	   \end{tabular}
\caption{{\em Upper panels:} dust density distribution at 2.5~Myr of evolution when a  20~$M_{\rm{Jup}}$ planet is embedded in the inner disk (10-15AU) and a second planet (15~$\rm{M_{\rm{Jup}}}$ {\em left panels}, 20~$\rm{M_{\rm{Jup}}}$  {\em right panels}) is injected in the outer region ($r_{p2}\gtrsim$70~AU) after 2~Myr of dust evolution. The white line is $\rm{St}=1$, and the blue line is $a_{\rm{frag}}$. {\em Bottom panels:} real part of the visibilities at $\lambda=[0.87 , 1.0, 3.0, 7.0]$mm at different times of dust evolution; ALMA Cycle 0 and ATCA data are over-plotted for comparison.}
   \label{two_planets_later}
\end{figure*}

In this case, a second trap also exists at the outer edge of the gap carved by the planet at $r_{p2}\sim$70~AU. 
The amount of dust concentrated in this second trap depends on the mass of the outer planet. 
When the planet is more massive, dust filtration becomes more effective and the millimetre flux increases. 
This has a significant influence on the resulting real part of the visibilities. 
Because of the dust trapping in both pressure maxima, the visibility profiles show strong oscillations. 
For the case of $M_{p2}=5~M_{\rm{Jup}}$, these oscillations are smoothed out late in the evolution ($\sim$10~Myr)
because of the continued crossing of dust particles through the outer gap, which is filling the inner trap 
(middle and bottom left panels of Fig.~\ref{two_planets_same_time}). 
In this case, there is a similar contrast of the inner to the outer ring emission as that observed with ALMA Cycle~0 (of the order of $\sim$100), 
generating similar visibility profiles to those observed. 
Nevertheless, for the case of a higher mass planet ($M_{p2}\gtrsim~10~M_{\rm{Jup}}$), 
the amount of dust concentrated in both traps is similar, 
and the undulating shape is more extreme and persists at late times, in contradiction to observations. 
The total flux at $\lambda=0.87$~mm is slightly lower than in the case of the 20~$\rm{M_{\rm{Jup}}}$ single planet; however, 
these fluxes can be higher if the disk mass or dust-to-gas ratio are assumed to be higher, or they can change by assuming different 
dust composition \citep{min2005}. 
We keep these two values the same for all the simulations to have a clear distinction of the effect of the planet 
parameters on the final dust distributions. 

From these two-planets simulations, in which both planets are considered to interact with the disk from early in the disk lifetime, 
we conclude that the outer planet should be a low-mass planet ($\lesssim~ 5~M_{\rm{Jup}}$) compared to the inner planet (20~$M_{\rm{Jup}}$),  
and the disk may be as old as in the single planet scenario ($\sim$5-10~Myr).

\subsubsection{Later injection of the outer planet}
High-contrast imaging of this disk with VLT/NACO shows signatures of a massive planet ($\sim15-20~M_{\rm Jup}$) 
in the outer region at $\sim70$~AU \citep{quanz2013, currie2014}. 
However, in the previous section we showed that if the planet in the outer disk is as old as the one in the inner disk, 
the resulting visibility profiles are in disagreement with observations. 
To have a very massive planet in the outer disk and similar millimetre emission contrast between the inner and the outer ring ($\sim100$), 
the outer planet {\em must} be younger than the inner planet. 
In this scenario, at the time that the outer planet is injected in the disk, most of the dust has already moved towards the inner trap, 
decreasing the mass of dust available for trapping in the second pressure bump.

\begin{figure}
 \centering
  	\includegraphics[width=8.5cm]{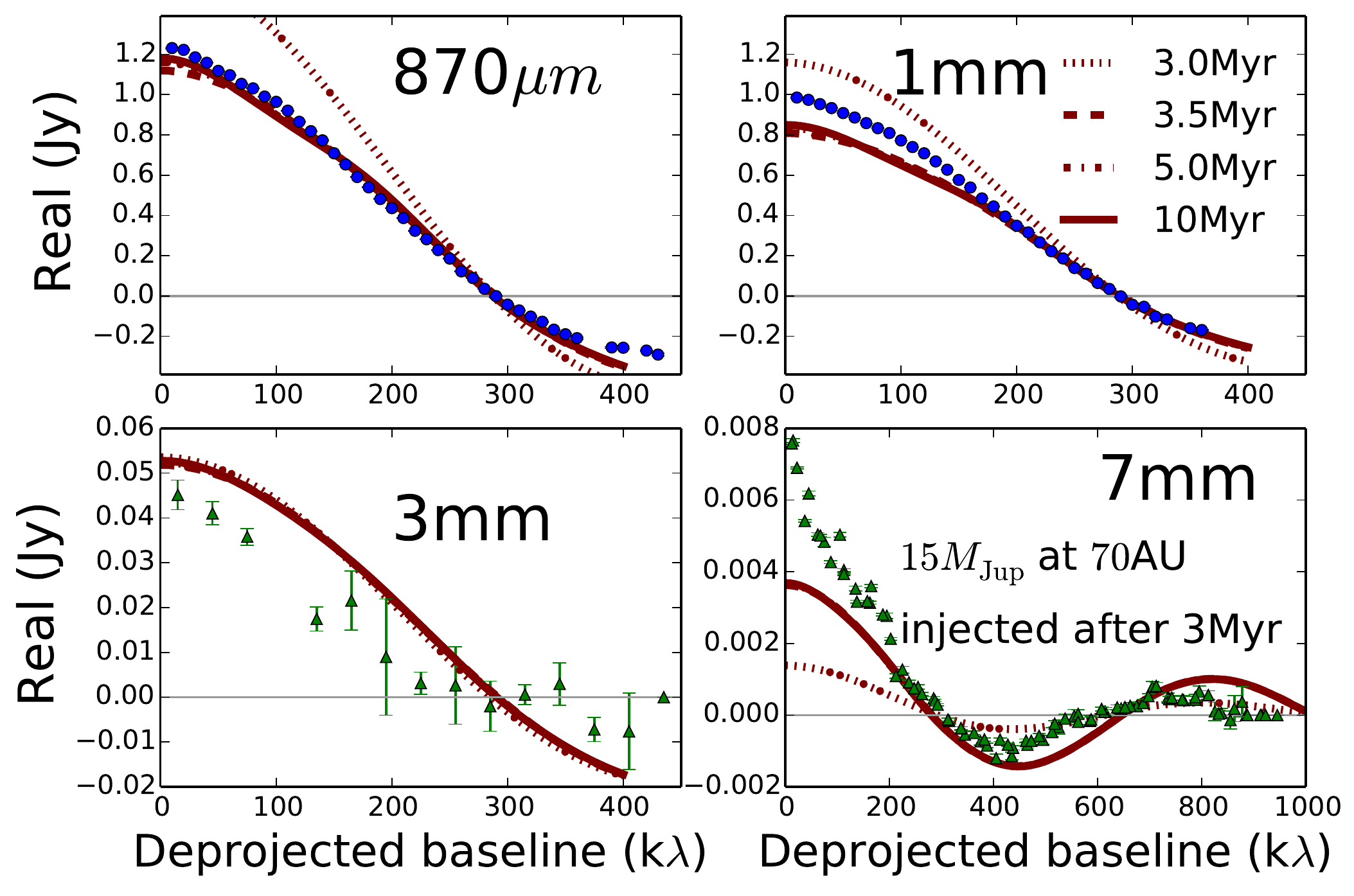}
\caption{Real part of the visibilities at $\lambda=[0.87 , 1.0, 3.0, 7.0]$mm at different times of dust evolution when a  20~$M_{\rm{Jup}}$ planet is embedded in the inner disk (10-15AU) from early stages, and a second planet (15~$\rm{M_{\rm{Jup}}}$) is injected after 3~Myr of dust evolution. ALMA Cycle 0 and ATCA data are over-plotted for comparison.}
   \label{after_3Myr}
\end{figure}

Figure~\ref{two_planets_later} shows the dust density distribution at 2.5~Myr of evolution when a 20~$M_{\rm{Jup}}$ planet 
is embedded in the inner disk, and a second planet ($M_{p2}=[15, 20]~M_{\rm{Jup}}$) is injected in the outer region 
($r_{p2}\sim$70~AU) after 2~Myr of dust evolution. 
In comparison with the corresponding cases of Fig.~\ref{two_planets_same_time}, 
it is important to notice that the outer ring is fainter in this case than when the two planets are assumed to interact 
with the disk over the same timescale. 
The oscillations on the visibility profiles are much smoother in these cases, and better fit the total flux. 
The null in the visibility profiles remains similar as in the single planet case ($\sim290~\rm{k}\lambda$). 
A 15~$M_{\rm{Jup}}$ outer planet fits the visibility shapes better and total fluxes than a 20~$M_{\rm{Jup}}$ planet, 
in agreement with the Gemini/NICI observations reported by \cite{currie2014}, 
who constrain the mass of the outer planet to be at most 15$M_{\rm{Jup}}$.  
The real part of the visibilities still have smooth oscillations; however, this can be further smoothed out 
by assuming than the planet is injected even later in the simulations.  
Figure~\ref{after_3Myr} displays the case of a 20~$M_{\rm{Jup}}$ planet embedded in the inner disk from early stages, 
and an outer 15~$\rm{M_{\rm{Jup}}}$ planet injected after 3~Myr of dust evolution.  
To reproduce the observations, the lower the mass of the planet, the earlier it needs to be introduced into the outer disk.
Therefore, assuming the mass predicted by \cite{quanz2013, quanz2014} and \cite{currie2014} for the outer planet 
(and its possible circumplanetary disk), implies that this planet must be {\em at least} $\gtrsim2-3$~Myr younger 
than the planet in the inner disk.

\section{Discussion}     \label{discussion}
We have shown that radial drift alone cannot reproduce the dust distribution observed for HD~100546 (Sect.~\ref{no_planet}). 
To obtain good agreement between the dust evolution models, which include radial drift, with the millimetre observations, 
an inner pressure bump \emph{must} exist since early in the disk lifetime ($<$0.1Myr).

Observations of HD~100546 indicate the presence of at least two companions in this disk \citep[see e.g.][]{acke2006, quanz2013}. 
The mass of the inner companion has largely been discussed by several authors \citep[e.g.][]{tatulli2011}. 
If the required inner pressure bump is induced by a planet, we demonstrated that the required planet must be 
massive ($\sim20~M_{\rm{Jup}}$) for a disk viscosity of $\alpha=2\times10^{-3}$. This
in agreement with \cite{mulders2013} results, but is derived from different observations (Sect~\ref{one_planet_section}). 
The trapping in HD~100546 disk is difficult because of the high disk temperature \citep[as derived in][]{bruderer2012} 
and disk viscosity, which both increase the relative motion of the particles due to turbulence, and decrease the maximum grain size 
that particles can reach under these assumptions. 
This maximum grain size is one order of magnitude lower than the size corresponding to $\rm{St}=1$, 
which are the particles that feel the highest radial drift and are the easiest to trap in a pressure bump. 
Assuming a higher viscosity would make the trapping even more difficult and a more massive inner planet, located closer to the star, 
would be needed to explain the millimetre emission. 
In the case of a single planet embedded in the inner part of the disk, the disk is more likely to be old ($\sim$5-10~Myr). 

In addition to the inner planet, we also assumed an outer planet as suggested by various authors e.g. \cite{boccaletti2013} and  \cite{quanz2013}. 
When the two planets are assumed to interact with the disk from the same stage of evolution, 
we demonstrated that the mass of the outer planet needs to be lower than the inner planet ($\lesssim~5~M_{\rm{Jup}}$), 
otherwise the resulting dust density distributions are in disagreement with millimetre observations. 
This is because, if a second massive planet is coeval with the inner planet, it traps inward drifting grains too effectively in the outer bump, 
and the contrast ratio between the bumps becomes too strong compared with observations. 
However, the mass suggested by various authors for the outer planet is higher than $5~M_{\rm{Jup}}$. 
As in the single planet case, the models are more consistent with observations when the disk is old ($\sim$~5-10Myr).  
To test how sensitive the resulting timescales are to the assumed gas surface density profile,  
we considered the best-fit single planet model and adopted an exponentially tapered disk rather than a power law. 
For this case, mm-grains are concentrated in a much narrower ring, shifting the null of the visibilities to longer baselines, 
in disagreement with observations. In addition, at very early times, there is an oscillating behaviour (similar to the previous two-planet simulations). 
Hence, with this gas surface density profile, stronger oscillating behaviour is expected for the visibilities for the case of two planets. 
Thus, a power law for the gas surface density gives us a better fit to the observations.

\begin{figure}
 \centering 
   	\includegraphics[width=8.5cm]{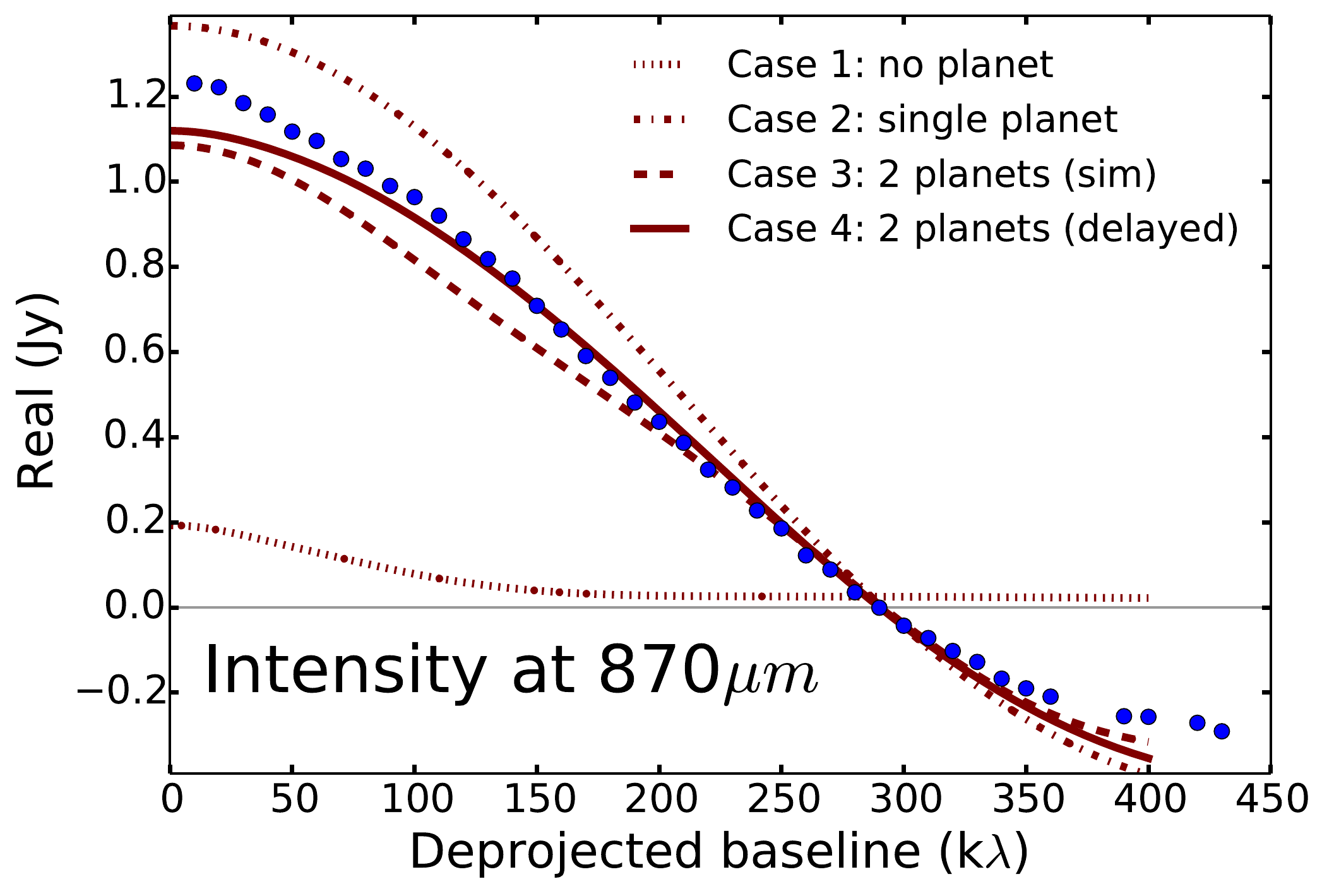}\\
   \caption{Real part of the visibilities at $870\mu$m. Case~1: no planet embedded in the disk and $t=2$~Myr. Case~2: a single planet embedded in the disk of 20$M_{\rm{Jup}}$ mass at $\sim$~10~AU and $t=10$~Myr. Case~3: two planets simultaneously interacting with the disk,  20$M_{\rm{Jup}}$ at $\sim$~10~AU and 5$M_{\rm{Jup}}$ at $\sim$~70~AU, and $t=10$~Myr. Case~4: two planets in the disk, but the outer planet is injected $t=3$Myr after the inner planet,   20$M_{\rm{Jup}}$ at $\sim$~10~AU and 15$M_{\rm{Jup}}$ at $\sim$~70~AU, and $t=3.5$~Myr. ALMA Cycle~0 is over-plotted for comparison.}
   \label{summary_visi}
\end{figure}

If the mass of the outer planet is assumed to be large, as suggested by \cite{quanz2013} ($\sim15~M_{\rm{Jup}}$), 
we demonstrated that the outer planet must be at least $\sim 2-3$~Myr younger than the inner planet, 
favouring a significantly younger outer planet observed in the act of formation. 
In this case, the disk can be much younger than when the disk only hosts a single planet.
 The required time to be in agreement with observations is a few tenths of a Myr after the outer planet is injected. 
 Figure~\ref{summary_visi} summarises our main findings and compares the cases of no planet, a single planet, 
 two planets injected simultaneously, and two planets where the outer planet is injected after 3~Myr of evolution. One of the uncertainties of these predictions is the disk midplane temperature, which may be slightly lower than obtained by \cite{bruderer2012} (Sect.~\ref{setup}). The temperature profile used is an upper limit; however, ALMA Cycle 0 CO J=3-2 observations show no signs of CO freeze-out, i.e. we also have a lower limit of $\sim$20 K.
A slightly lower midplane temperature decreases the dust drift velocities (Eq~\ref{dustvel}), and increase the maximum grain size (Eqs.~\ref{adrift} and \ref{afrag}). Observations with higher angular resolution and sensitivity of optically thin emission from multiple transitions of CO isotopologues are needed to give better constraints on the disk midplane temperature and confirm the current predictions.  However, we expect that the overall trends remain similar with a slight lowering of the dust temperature.

\begin{figure}
\centering
 	\includegraphics[width=8.5cm]{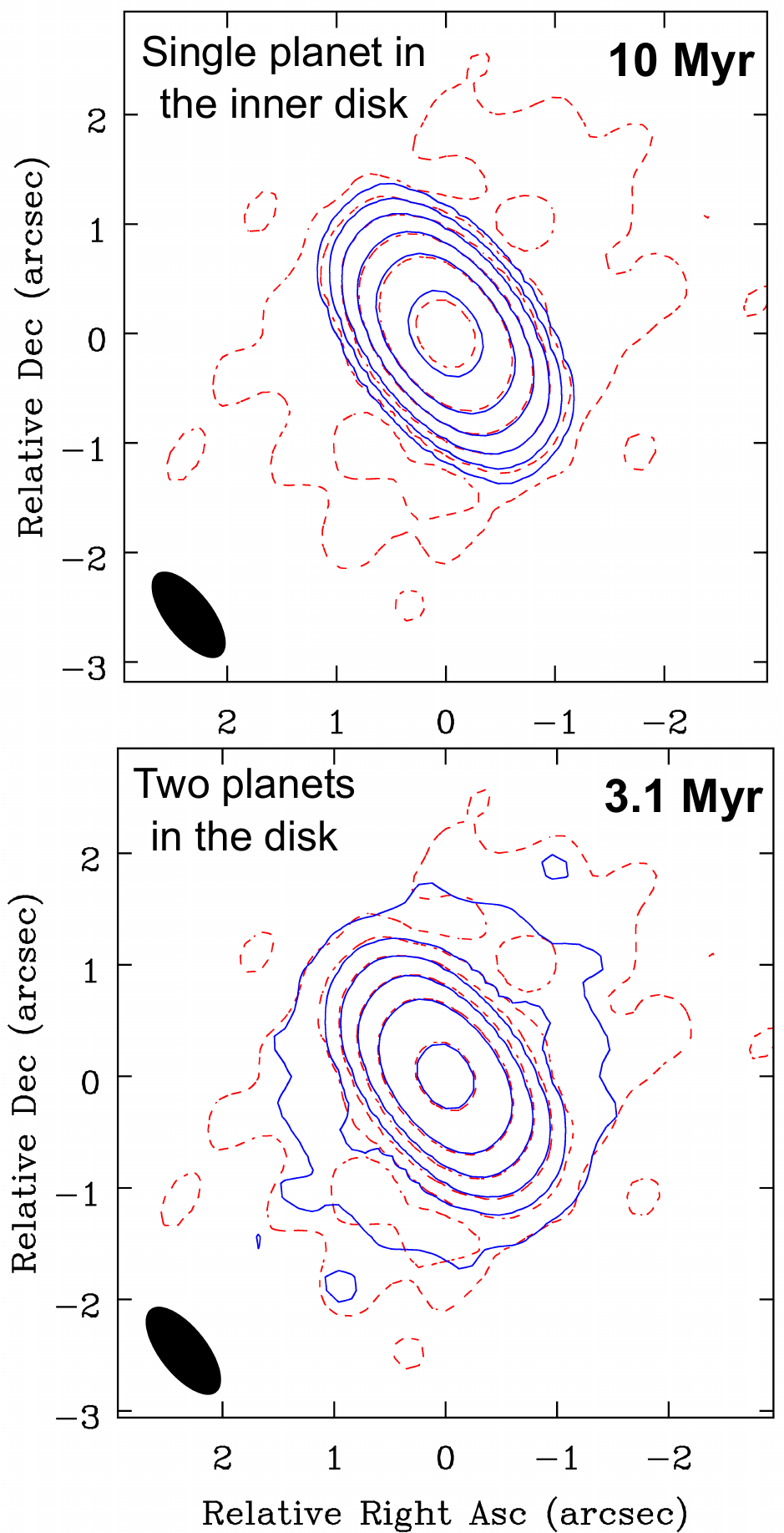}
\caption{ALMA Cycle 0 continuum contours at 870$\mu$m (red dashed lines) and the contours corresponding to the models (blue solid lines) when an inner  20 $\rm{M_{\rm{Jup}}}$ planet alone ({\em top panel}) is embedded in the disk, and with two planets in the disk ({\em bottom panel}), where the outer planet is injected after 3~Myr of evolution.  Contours are every 3, 10, 30, 100, 300, and 1000 times the rms (0.5 mJy beam$^{-1}$).}
  \label{alma_model_obs}
\end{figure}

\begin{figure}
\centering
 	\includegraphics[width=8.5cm]{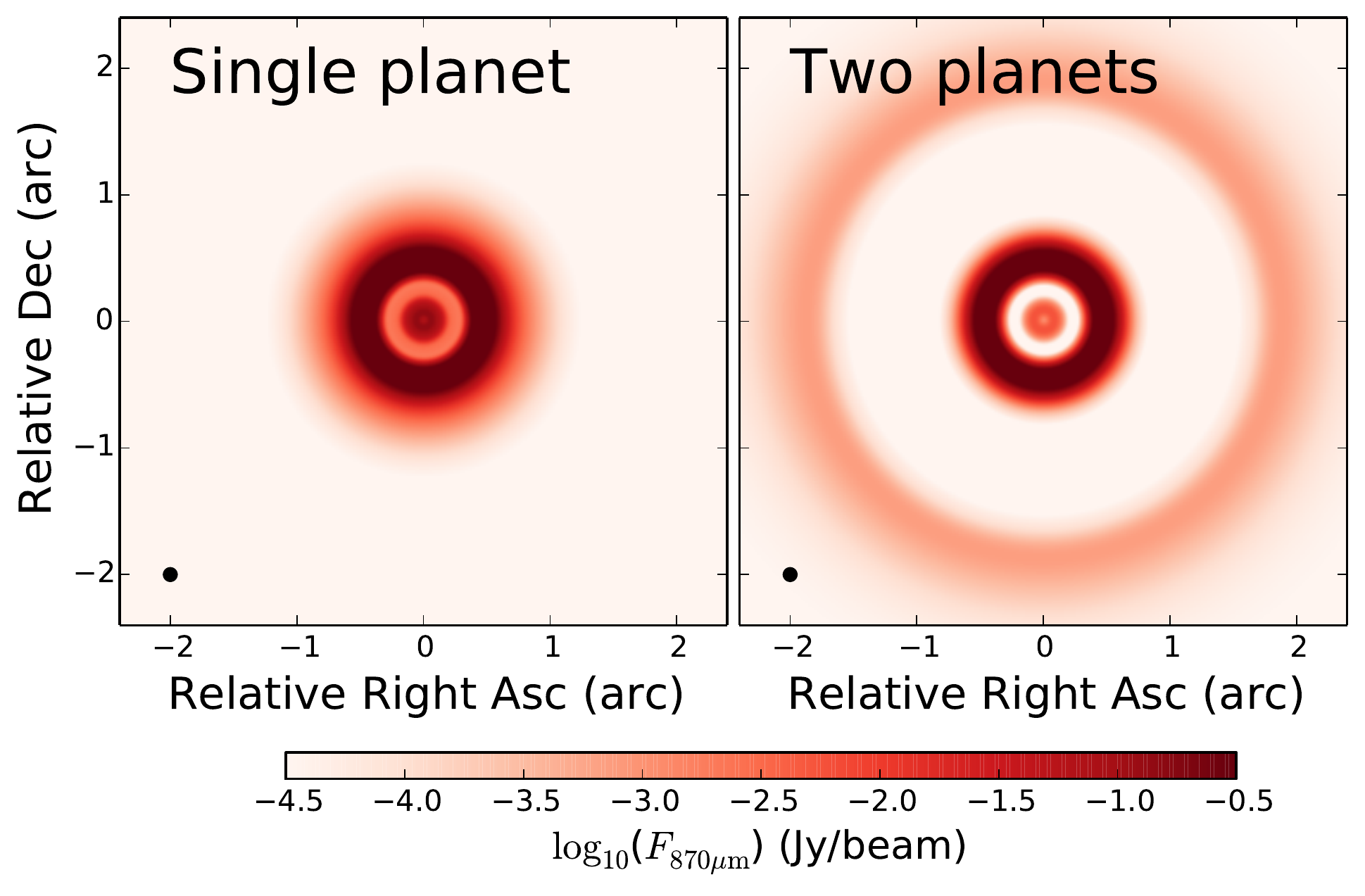}
\caption{Synthetic images at 870~$\mu$m convolved with a $0.1''\times 0.1''$ beam from the models of a single planet (20~$M_{\rm{Jup}}$ at $\sim$10~AU) and two planets (20~$M_{\rm{Jup}}$ at $\sim$10~AU and 15~$M_{\rm{Jup}}$ at $\sim$70~AU), injected 3~Myr after the inner planet.}
 \label{future_obs}
\end{figure}

Moreover, Fig.~\ref{alma_model_obs} shows the synthetic ALMA Cycle 0 contour lines image at 870$\mu$m 
(imaged using identical ($u$, $v$) coordinates as the ALMA observations), 
for the cases of one single planet and an old disk, and two massive planets and a younger disk. 
The latter scenario reproduces a more extended emission than with a single planet, but it is not as extended as observed. 
This discrepancy is because the width of the carved gaps (thus the location of the pressure maximum at the outer edge of the gap) 
is underestimated in our models compared to proper hydrodynamical simulations (Sect~\ref{carved_gaps}). 
The location of the pressure maximum of a gap carved by a $15~M_{\rm{Jup}}$ planet at 70~AU is expected to be 
$\sim~190$~AU i.e. at $9-10 r_H$ from the planet location, where the peak of the outer ring emission is observed. 
Figure~\ref{future_obs} illustrates the synthetic images at 870~$\mu$m convolved with a $0.1''\times 0.1''$ beam from the same 
models used for Fig~\ref{alma_model_obs}, showing that with high angular resolution observations, two rings can be resolved in the two-planet scenario.

The age of  HD~100546 itself is very uncertain \citep[$\sim$3-10Myr][]{acke2006} and a more precise measurement of the age of 
this star can give us hints as to the nature of the potential planets embedded in this disk. 
An observation to ultimately test dust trapping in two pressure bumps, would be to measure spectral index 
variations inside and outside these pressure maxima locations \citep{pinilla2014a}, which are testable with 
future ALMA capabilities. 
Figure~\ref{spectral_index} shows the expected radial variations of the spectral index, calculated between 1 and 3~mm, 
when a single inner planet of 20~$M_{\rm{Jup}}$ mass is embedded in the inner disk at $\sim$10~AU, 
and when two planets are in the disk, but the outer planet (15~$M_{\rm{Jup}}$) is injected 3Myr after the inner planet 
(20$M_{\rm{Jup}}$). 
At the location of the pressure trap, the spectral index is lower because of the accumulation of mm-grains and vice versa. 

We also used ATCA observations to compare the visibility profiles at 3 and 7~mm with the models. 
The models with two planets predict slightly better the total flux, and estimate well the observed null of the real part of the visibilities 
at these wavelengths. 
The null at 7~mm is at a moderately longer $k\lambda$ than at shorter wavelengths. 
This happens because larger grains are expected to be more concentrated. 
Nonetheless, for all cases, the total flux at 7mm is under-predicted, 
which partially comes from the neglect of free-free emission in the models, and because of the small maximum grain size ($\sim$~1mm) 
under the assumed temperature and disk viscosity in the outer disk. 

\section{Conclusion} \label {conclusion}

\begin{figure}
 \centering
  	\includegraphics[width=8.5cm]{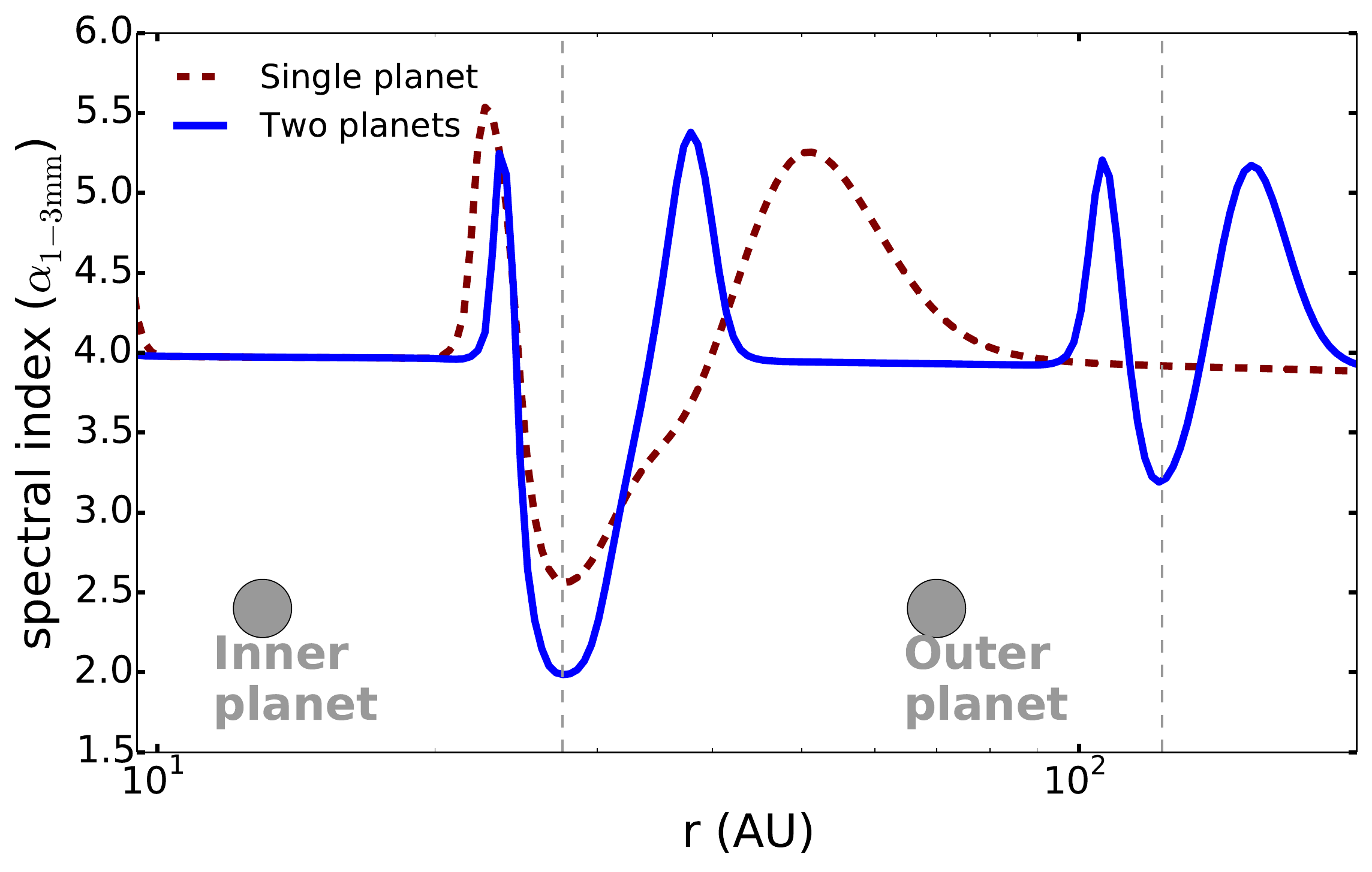}
\caption{Radial variations of the spectral index, calculated between 1 and 3~mm, when a single inner planet of  20~$M_{\rm{Jup}}$ mass is embedded in the disks and when two planets are in the disk, but the outer planet (15~$M_{\rm{Jup}}$) is injected 3Myr after the inner planet (20$M_{\rm{Jup}}$). The vertical lines correspond to the positions of the pressure maxima at the outer edge of the gap(s) carved by the planet(s).}
   \label{spectral_index}
\end{figure}

To obtain good agreement between dust evolution models and millimetre observations of the disk around HD~100546, 
an inner pressure trap must exist from early in the disk lifetime ($\lesssim$1~Myr). 
If this pressure bump is formed because of a planet embedded in the inner disk ($r\sim$10~AU), 
the mass of the planet should be high $(\sim 20~M_{\rm{Jup}})$ for a disk viscosity of $\alpha=2\times10^{-3}$, 
in agreement with \cite{mulders2013}, who found similar results from modelling mid-infrared data. 
In the case in which this is the only planet embedded in the disk, the disk is more likely to be old ($\sim$5-10~Myr). 
In this case the outer ring of emission observed with ALMA remains unexplained (Fig~\ref{alma_model_obs}). 
If an outer planet is also embedded in the disk, and it is as massive as suggested by \cite{quanz2013, quanz2014} and \cite{currie2014} ($\sim15~M_{\rm{Jup}}$), 
this outer planet should be at least $\sim 2-3$Myr younger than the inner planet, supporting the hypothesis that the outer planet may be in the 
act of formation. 
In this case, the models produce an outer ring of emission that is 100 times fainter than the inner ring of emission as observed with ALMA Cycle 0 
(Fig~\ref{alma_model_obs}).  
If the outer planet is embedded in the disk at the same time as the inner planet, inward drifting grains are efficiently trapped in the outer bump, 
and the contrast of the millimetre fluxes between the two pressure traps becomes too strong compared with the observations. The
ATCA observations at 3 and 7~mm also favour the two-planets scenario. Future high angular resolution and sensitivity observations of multiple transitions of optically thin emission from CO isotopologues, which can constrain better the disk gas surface density and temperature profiles, and  continuum images with ALMA will allow us to 
resolve potential rings (Fig.~\ref{future_obs}), to measure spectral index variations inside and outside pressure maxima locations 
(Fig.~\ref{spectral_index}), and to confirm particle trapping by one or two planets in this disk.

\begin{acknowledgements}
The authors are very grateful to E.~F.~van~Dishoeck and D.~Harsono for all their comments and fruitful discussions. 
We thank to S.~Bruderer for providing his data for the disk temperature and C.~Wright for the ATCA data. 
P.~P. is supported by Koninklijke Nederlandse Akademie van Wetenschappen (KNAW) professor prize to Ewine van Dishoeck. 
T.~B. acknowledges support from NASA Origins of Solar Systems grant NNX12AJ04G. 
C.~W. acknowledges support from the Netherlands Organisation for Scientific Research (NWO, grant number 639.041.335). 
Astrochemistry in Leiden is supported by the Netherlands Research School for Astronomy (NOVA), 
by a Royal Netherlands Academy of Arts and Sciences (KNAW) professor prize, 
and by the European Union A-ERC grant 291141 CHEMPLAN.  
This paper makes use of the following ALMA data: ADS/JAO.ALMA\#2011.0.00863.S. 
ALMA is a partnership of ESO (representing its member states), NSF (USA) and NINS (Japan), together with NRC (Canada) and NSC and ASIAA (Taiwan), 
in cooperation with the Republic of Chile.  The Joint ALMA Observatory is operated by ESO, AUI/NRAO and NAOJ.  

\end{acknowledgements}

\bibliographystyle{aa}

\bibliography{hd100546_pinilla.bbl}

\end{document}